\documentclass[arguments]{aastex631}

\usepackage{booktabs}
\usepackage{array,multirow}
\usepackage{tabularx}
\usepackage{rotating}
\usepackage{amsmath}

\shorttitle{NEOs by ATLAS}
\shortauthors{Deienno et al.}
\graphicspath{{./}{figures/}}

\begin{document}

\title{The Debiased Near-Earth Object Population from ATLAS Telescopes}

\correspondingauthor{Rogerio Deienno}
\email{rogerio.deienno@swri.org, rdeienno@boulder.swri.edu}

\author[0000-0001-6730-7857]{Rogerio Deienno}
\affiliation{Solar System Science \& Exploration Division, Southwest Research Institute, 1301 Walnut Street, Suite 400, Boulder, CO 80302, USA}

\author[0000-0002-7034-148X]{Larry Denneau}
\affiliation{Institute for Astronomy, University of Hawaii, 2680 Woodlawn, Honolulu, HI, 96822, USA}

\author[0000-0002-4547-4301]{David Nesvorn\'y}
\affiliation{Solar System Science \& Exploration Division, Southwest Research Institute, 1301 Walnut Street, Suite 400, Boulder, CO 80302, USA}

\author[0000-0002-6034-5452]{David Vokrouhlick\'y}
\affiliation{Institute of Astronomy, Charles University, V Hole\v{s}ovi\v{c}k\'ach 2, CZ–18000, Prague 8, Czech Republic}

\author[0000-0002-1804-7814]{William F. Bottke}
\affiliation{Solar System Science \& Exploration Division, Southwest Research Institute, 1301 Walnut Street, Suite 400, Boulder, CO 80302, USA}

\author[0000-0001-7830-028X]{Robert Jedicke}
\affiliation{Institute for Astronomy, University of Hawaii, 2680 Woodlawn, Honolulu, HI, 96822, USA}

\author[0000-0003-4439-7014]{Shantanu Naidu}
\affiliation{Jet Propulsion Laboratory, California Institute of Technology, 4800 Oak Grove Dr. Pasadena, CA 91109, USA}

\author[0000-0003-3240-6497]{Steven R. Chesley}
\affiliation{Jet Propulsion Laboratory, California Institute of Technology, 4800 Oak Grove Dr. Pasadena, CA 91109, USA}

\author[0000-0003-0774-884X]{Davide Farnocchia}
\affiliation{Jet Propulsion Laboratory, California Institute of Technology, 4800 Oak Grove Dr. Pasadena, CA 91109, USA}

\author[0000-0002-6381-8534]{Paul W. Chodas}
\affiliation{Jet Propulsion Laboratory, California Institute of Technology, 4800 Oak Grove Dr. Pasadena, CA 91109, USA}




\begin{abstract}
This work is dedicated to debias the Near-Earth Objects (NEO) population based on observations from the Asteroid Terrestrial-impact Last Alert System (ATLAS) telescopes. We have applied similar methods used to develop the recently released NEO model generator (NEOMOD), once debiasing the NEO population using data from Catalina Sky Survey (CSS) G96 telescope. ATLAS is composed of four different telescopes. We first analyzed observational data from each of all four telescopes separately and later combined them. Our results highlight main differences between CSS and ATLAS, e.g., sky coverage and survey power at debiasing the NEO population. ATLAS has a much larger sky coverage than CSS, allowing it to find bright NEOs that would be constantly ``hiding'' from CSS. Consequently, ATLAS is more powerful than CSS at debiasing the NEO population for H $\lesssim$ 19. With its intrinsically greater sensitivity and emphasis on observing near opposition, CSS excels in the debiasing of smaller objects. ATLAS, as an all sky survey designed to find imminent hazardous objects, necessarily spends a significant fraction of time looking at places on the sky where objects do not appear, reducing its power for debiasing the population of small objects. We estimate a NEO population completeness of $\approx$ 88\%$^{+3\%}_{-2\%}$ for H $<$ 17.75 and $\approx$ 36\%$^{+1\%}_{-1\%}$ for H $<$ 22.25. Those numbers are similar to previous estimates (within error bars for H $<$ 17.75) from CSS, yet, around 3\% and 8\% smaller at their face values, respectively. We also confirm previous finding that the $\nu_6$ secular resonance is the main source of small and faint NEOs at H = 28, whereas the 3:1 mean motion resonance with Jupiter dominates for larger and brighter NEOs at H = 15.
\end{abstract}



\section{Introduction} \label{sec:intro}

Near-Earth Objects (NEOs) are asteroids and comets whose orbits come close to or cross Earth's  orbit (perihelion distance q $<$ 1.3 au), and as such may pose an impact hazard to Earth. Finding, tracking, and understanding NEOs are among the main objectives of the NASA's Planetary Defense Coordination Office (PDCO)\footnote{\url{https://science.nasa.gov/planetary-defense}}. Particularly, estimating the NEO population at several different absolute magnitude (H) ranges is crucial to understanding the completeness of the catalog of known objects (i.e., completeness is a measure of the fraction of the NEO population yet to be found). 

An efficient way to estimate the NEO population, and its completeness, is by debiasing 
 NEO surveys. The goal is to determine how efficient a given survey's telescope is at observing NEOs of different H values by comparing those numbers with real NEOs detected by the same survey's telescope. Considerable effort has gone to debiasing sky surveys over the last several years \citep[e.g.][]{Tricarico2016,Tricarico2017,Granvik2018,Harris2021,Harris2023}. Our work in this paper builds on efforts from  \citet{Nesvorny2023neomod,Nesvorny2024neomod2} to debias NEO detections from the Mt. Lemmon (observatory code G96) telescope of the Catalina Sky Survey\footnote{\url{https://catalina.lpl.arizona.edu}} \citep[CSS;][]{Christensen2012}. The result of their modeling work was named NEOMOD Simulator 1 and 2, hereafter referred to as NEOMOD1 and NEOMOD2. They are powerful tools capable of generating debiased orbital and magnitude distributions for NEOs between 15 $\leq$ H $\leq$ 28 as desired and defined by the user.

The development of NEOMOD1 took into consideration observations from G96 and the Catalina telescope on Mt. Bigelow near Mt. Lemmon (observatory code 703) between the years 2005 and 2012 \cite[included;][]{Nesvorny2023neomod}. Field of View (FoV) pointing and the survey telescopes efficiencies over the aforementioned period were obtained from \citet{Jedicke2016}. The model of \citet{Nesvorny2023neomod} was then compared with findings by \citet{Granvik2018}, which had previously used the same data set as provided by \citet{Jedicke2016}. The advantage of the work by \citet{Nesvorny2023neomod} was that it used a new and more accurate methodology for debiasing sky surveys. It took advantage of the publicly available {\tt objectsInField}\footnote{\url{https://github.com/AsteroidSurveySimulator/objectsInField}} code ({\tt oIF}) from the {\tt Asteroid Survey Simulator} ({\tt AstSim}) package \citep{Naidu2017} to simulate enormous sets of synthetic NEOs  (N $>$ 10$^8$ objects total) and record those that were potential candidates for being observed by our target sky survey. Combining results from {\tt oIF} with the respective telescope efficiencies, \citet{Nesvorny2023neomod} was able to debias observations from the CSS G96 and 703 telescopes over all NEO space (i.e., determine the fraction of NEOs that are expected to be detected as a function of binned semimajor axis ($a$), eccentricity ($e$), inclination ($i$), and H -- the model bias function $\mathcal{M}_b$($a,e,i,$H)). The methods used by \citet{Nesvorny2023neomod} also accounted for numerically simulating a large number (1.1$\times$10$^{6}$) of asteroids escaping from eleven different source regions within the main belt (10$^5$ objects per source) while also accounting for comets as the twelfth source \citep[not simulated but taken from][]{Nesvorny2017comets}. The \citet{Nesvorny2023neomod} model also self-consistently accounted for the disruption of NEOs at small perihelion distances \citep{Granvik2016}. 

The {\tt MultiNest} code \citep{Feroz2008,Feroz2009} was used to optimize the model fit to CSS detections. In simple terms (details in Section \ref{sec:model}), {\tt MultiNest} crosschecks information from the generated model bias function $\mathcal{M}_b$ with real unique detections of NEOs from CSS telescopes (i.e. not accounting for re-detections). With this information, it calculates maximum likelihood values to estimate the number of NEOs in each $a,e,i$,H bin, with the results yielding a probability density function (PDF) that includes correlations. These PDFs are then normalized to 1. 

The NEOMOD algorithm is based on millions of fit-trials performed by {\tt MultiNest}. The highest likelihood values for the PDF are used to generate the debiased population of NEOs, namely their orbital and magnitude distributions along with their respective 3$\sigma$ deviations. Such data can ultimately be used for generating a debiased cumulative magnitude distribution of NEOs. When compared to the population of known NEOs from the Minor Planet Center (MPC) catalog\footnote{\url{https://minorplanetcenter.net//iau/MPCORB.html}}, the result can be used to determine the completeness of the NEO population for different H ranges. NEOMOD1 was optimized to operate in the magnitude range 15 $\leq$ H $\leq$ 25. 

NEOMOD2 is an update of NEOMOD1 that was improved to cover the extended magnitude range 15 $\leq$ H $\leq$ 28. NEOMOD2 was developed using the same methods as NEOMOD1 but it covered CSS observations between 2013 and 2022 for the Mt. Lemmon G96 telescope. During this extended period, G96 detected over an order of magnitude more NEOs than those detected between 2005 to 2012. Many of the new detections were also fainter, with H $>$ 25. The differences were mostly the result of an upgrade in the G96 camera that went on-line in May 2016. Here the survey's FoV changed from 1.1$\degr \times$ 1.1$\degr$ to 2.23$\degr \times$ 2.23$\degr$ \citep[i.e., a factor of about 4.11 increase in square degrees FoV area; see Figure 1 in][]{Nesvorny2024neomod2}. Using these data, \citet{Nesvorny2024neomod2} were able to better characterize how loss effects due to fast moving objects interfere with the telescope detection efficiency (i.e., trailing effects; see Section \ref{sec:trlloss}). Losses from fast moving objects mostly affect the detection of faint objects that can only be detected once they are very close to Earth.  Examples would include those with H $\gtrsim$ 22, given that the G96 limiting detectable visual magnitude is V$_{\rm lim}^{\rm G96} \approx$ 22. Altogether, the large number of new detections, combined with better characterized detection efficiencies, allowed \citet{Nesvorny2024neomod2} to derive a reasonable bias function and fit for the NEO distribution beyond H = 25, the limiting magnitude of NEOMOD1. Note that the work by \citet{Nesvorny2024neomod2} did not include the Mt. Bigelow 703 observations because this telescope did not experience a substantial increase in the number of unique detections. The reason is that this telescope only had a lower limiting visual magnitude close to V = 19 mag, and therefore G96 detections dominate the time period in question. 

The results of NEOMOD2, and its debiased population, were compared against the findings of previous efforts from \citet{Harris2021}. This paper updated the work of \citet{Harris2015}, which for years was the base reference for our understanding of the NEO population. Unlike \citet{Nesvorny2024neomod2}, the work by \citet{Harris2021} is based on re-detections of NEOs from all NEO surveys. Despite the different methods applied in \citet{Nesvorny2024neomod2} and \citet{Harris2021} for debiasing the NEO population, their results are in reasonable agreement for objects with H $\lesssim$ 24. For H $>$ 24 \citet{Nesvorny2024neomod2} predicted a lower cumulative number of expected NEOs. This implies that the known NEO population is more complete than previously predicted by \citet{Harris2021}. 

The re-detection model used by \citet{Harris2021} is probably accurate up to H $\approx$ 24 \citep{Harris2015} because the number of new and re-detected NEOs is statistically large. For fainter magnitudes, \citet{Harris2021} could only rely on limited statistics. That is because they considered a `small' set of 100,000 orbits and when extending to expected completion levels below 0.001, statistics become compromised due to too few detections. As a work-around, they compared their results with what was known of bolides striking the Earth over time \citep{Brown2002,Brown2013}, but did not base their prediction on that rate. \citet{Harris2021} switched to an assumed slope for the completion curve, rather than one based on discovery statistics, or constant S/N of detections. This allowed them to extrapolate their estimate of the cumulative number of NEOs with H $\approx$ 24 to fainter (and smaller) objects. As discussed in \citet{Nesvorny2024neomod2}, this estimate produces a cumulative distribution that is steeper than that predicted by NEOMOD2. As a bottom line, one should not take the estimate from \citet{Harris2021} as face-value for H $>$ 24 or so, as it is based on an assumed completion model. Yet, an intriguing way to explain this difference between the estimate by \citet{Nesvorny2024neomod2} and \citet{Harris2021} is by considering that many small NEOs (and bolides) are a consequence of modestly-large NEOs tidally disrupting near Earth \citep{Granvik2024} (see also \citet{Nesvorny2024neomod2}.

The work by \citet{Harris2021} has been revised in \citet{Harris2023} taking into account a systematic positive offset in H magnitudes from MPC catalog \citep[e.g.,][]{Pravec2012}. Overall, the new absolute magnitudes reported in \citet{Harris2023} are generally fainter than what was used in \citet{Harris2021}. As the calibration of 
photometric observations improves over time, the absolute magnitudes of NEOs in the MPC catalog are revised. This effect appears to be systematic in that the H values of a population of faint NEOs increased (on average) by a fraction of H magnitude. \citet{Harris2023} updated NEO population shows a decrease of about 10\% at larger sizes from previous estimate by \citet{Harris2021}. As explained in \citet{Nesvorny2024neomod2}, however, this should not largely affect the completeness percentage because the revised absolute magnitudes should affect in the same way both the known and estimated population, i.e. their fraction remains roughly the same. Yet, \citet{Nesvorny2024neomod2} found the population of bright NEOs (H $<$ 17.75) to be about 91\%$_{\rm -4\%}^{\rm +4\%}$ complete, or about 4\% below what previously estimated by \citet{Harris2021}. Nonetheless, \citet{Nesvorny2024neomod2} also acknowledged that because their method for debiasing the NEO population relies on unique (single) detections, and the number of bright NEOs is relatively small, the completeness estimate of bright NEOs by \citet{Harris2021,Harris2023} could be more accurate due to larger statistics provided by the re-detection method.

More recently, a third version of the NEOMOD Simulator was released \citep[NEOMOD3;][]{Nesvorny2024neomod3}\footnote{\url{https://www.boulder.swri.edu/\~davidn/NEOMOD_Simulator}}. This new update includes additional constraints, namely several hundred asteroid albedos from the Wide-Field Infrared Survey Explorer\footnote{\url{https://www.jpl.nasa.gov/missions/wide-field-infrared-survey-explorer-wise}} (WISE). In NEOMOD3, \citet{Nesvorny2024neomod3} combined albedo and diameter measurements from the cryogenic portion of the NEOWISE mission \citep[see also][]{Mainzer2011} with the expected absolute magnitude distributions from NEOMOD2 to obtain the expected (debiased) size-frequency distribution of NEOs. Then, using {\tt MultiNest}, they determined the best fit log-likelihood value across all model parameters and size-frequency distribution possibilities \citep[see details in][]{Nesvorny2024neomod3}. This approach was more advantageous than choosing a reference albedo from NEOWISE asteroid data and then using it to translate the cumulative magnitude distribution into a cumulative size distribution \citep[e.g.,][]{Harris2021}, especially because NEOs have a wide range of visible albedos \citep[see also \citet{Morbidelli2020}]{Mainzer2011}.

In this work, we devote our efforts to debiasing the NEO population based on observations by the Asteroid Terrestrial-impact Last Alert System\footnote{\url{https://atlas.fallingstar.com}} (ATLAS). ATLAS is composed of four telescopes, two in the Northern hemisphere and two recent additions in the Southern hemisphere. Those in the North are both located in Hawaii; ATLAS-MLO is on Mauna Loa (observatory code T08) and ATLAS-HKO is on Haleakal\=a (observatory code T05). The two telescopes in the south are ATLAS South Africa, located in Sutherland Observing Station in South Africa (observatory code M22), and ATLAS Chile (observatory code W68),in Rio Hurtado. Hereafter we refer to those sites/telescopes as Mauna Loa (T08), Haleakal\=a (T05), Sutherland (M22), and Chile (W68).

The works by \citet{Tonry_2018} and \citet{Heinze2021} are the most recent updates on ATLAS characteristics and performance. The former describes the capabilities and operation procedures for ATLAS, while the latter is dedicated to understanding its capabilities. For example, \citet{Heinze2021} is the first to attempt debiasing the NEO population using ATLAS observations. The methods discussed in \citet{Heinze2021} differ from those used by \citet{Nesvorny2023neomod,Nesvorny2024neomod2} when developing NEOMOD 1 and 2. 

As detailed in their Section 2.2, \citet{Heinze2021} procedure consists of the following steps.  First, they simulate a large number of fictitious NEO orbits based on the distribution provided by \citet{Granvik2018} for 15 $<$ H $<$ 25. Second,  they `painted' tracklets (trail lengths) of fictitious NEOs into images for those orbits that overlapped with the ATLAS FoVs (as actual NEOs would produce in real ATLAS images). Third, they applied the ATLAS detection pipeline \citep{Tonry_2018} from the Moving Object Processing System \citep[MOPS;][]{Denneau2013} technique to assess whether the fictitious NEO would be detected. Using this procedure, \citet{Heinze2021} was able of estimate the detection fraction of NEOs $f_d(H)$ as a function of H. Thus, by comparing $f_d(H)$ with the number of real ATLAS detections, they were able to estimate the nature of the debiased NEO population.

Our methods to debias ATLAS telescopes are similar to those presented by \citet{Nesvorny2023neomod,Nesvorny2024neomod2}.  Specifically, we simulated a very large number of synthetic NEO orbits with {\tt oIF} \citep{Naidu2017} to determine their overlap with each of four ATLAS telescopes' FoVs, and then applied those ATLAS telescopes' derived efficiencies (see Section \ref{sec:model} for details) to derive their individual model biases ($\mathcal{M}_b$). It is important that we compare our results with those from \citet{Heinze2021}, partly because we can confirm that the methods used by  \citet{Nesvorny2023neomod,Nesvorny2024neomod2} are flexible enough to be applied to different surveys, but also because it allows us to calibrate our results with previous work. 

The one clear advantage that we have over \citet{Heinze2021} is that we have access to more NEO detections. \citet{Heinze2021} covered a one year period for the ATLAS Northern telescopes (from 2017 June 01 through 2018 August 22), in which ATLAS detected 713 distinct NEOs (unique detections). The ATLAS Sutherland (M22) and Chile (W68) telescopes were not yet operational.  In this work, we use 8 years of ATLAS operations, where ATLAS detected 4646 distinct NEOs. These data include detections from over a year of work from the ATLAS Sutherland (M22) and Chile (W68) telescopes.  The detection numbers are as follows: Mauna Loa (T08 -- period: 2017 - 2022) detected 3482 distinct NEOs, Haleakal\=a (T05 -- period: 2015 - 2023) detected 3488 distinct NEOs, Sutherland (M22) and Chile (W68) detected 700 and 812 distinct NEOs, respectively, between the years 2022 and 2023. Note that a large fraction of distinct/unique NEO detections were made by two or more ATLAS telescopes. The duplicates are removed, explaining why our count of 4646 distinct NEOs detections is smaller than the sum of the values above.

\begin{figure}[h!]
    \centering
    \includegraphics[width=1.\linewidth]{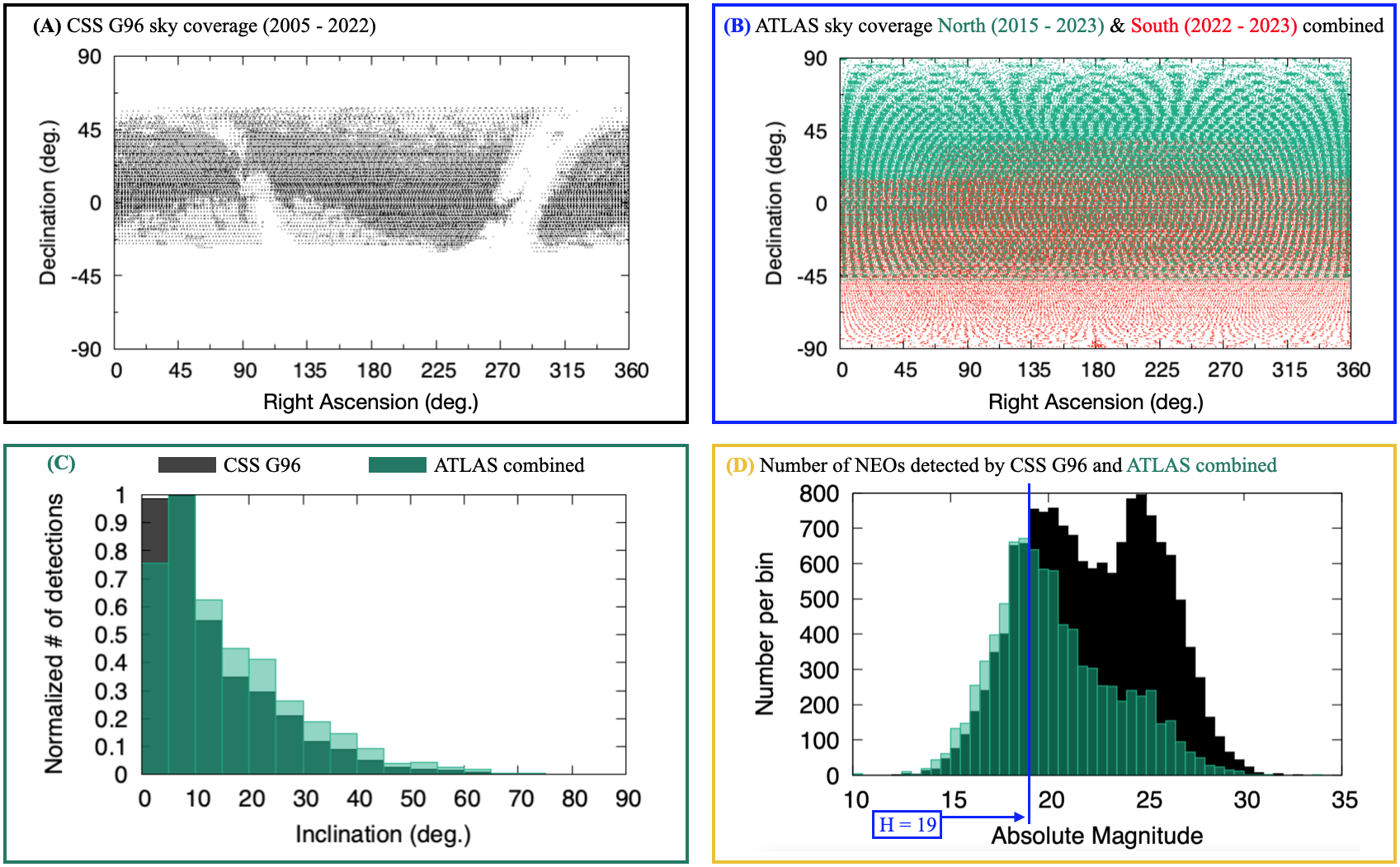}
    \caption{Comparison between CSS G96 and ATLAS telescopes. (A - black box): CSS G96 sky coverage between the years 2005 and 2022. The white stripes in this figure shows the location of the galactic plane, where G96 avoids observing. (B - blue box): ATLAS sky coverage combining two Northern (green dots) telescopes, i.e. Mauna Loa (T08 -- period: 2017 - 2022) and Haleakal\=a (T05 -- period: 2015 - 2023), as well as two Southern (red dots) telescopes, i.e. Sutherland and Chile (M22 and W68, respectively -- period: 2022 - 2023 for both). (C - green box): Normalized number of uniquely detected objects as a function of orbital inclination. This figure shows all unique detections made by CSS G96 (black) and by ATLAS when combining all its 4 telescopes (green), over the same periods reported in panels A and B. (D - yellow box): Distribution of NEOs unique detections made by CSS G96 (black) and by ATLAS when combining all its 4 telescopes (green) as a function of absolute magnitude (period covered as described in panels A, B, and C). The blue line at label mark H = 19 indicates the transition in absolute magnitude where ATLAS has greater debiasing power than CSS (H $<$ 19), and vice-versa. The excess of objects in the CSS detections at $H=25$ is due to a recent improvement in the CSS G96 detector \citep[see Fig. 1 in][]{Nesvorny2024neomod2}.}
    \label{Fig1}
\end{figure}

In this work, we also want to understand how the NEO population, as debiased from ATLAS observations, compares to that of NEOMOD2 (CSS G96) as well as that of \citet{Harris2021}. We are particularly interested in comparisons between these works in terms of population completeness. We do not compare our results directly with \citet{Harris2023} as those were not yet published in a final paper format and mostly because, in terms of population completeness, they should be similar to the findings published in \citet{Harris2021}. Although the G96 dataset used for developing NEOMOD2 is large enough to obtain accurate results, it is still important to compare the NEOMOD2 population and completeness predictions with those from ATLAS because the latter has very different capabilities from G96. For example, in some ways, ATLAS is less capable; it has a limiting magnitude sensitivity of V$_{\rm lim}^{\rm ATLAS} \approx$ 19.7, which is not as powerful as G96's V$_{\rm lim}^{\rm G96} \approx$ 22.  In other respects, however, ATLAS has several advantages, as described below and as shown in Figure \ref{Fig1}:
\begin{itemize}
    \item ATLAS full sky coverage is much broader than G96 (Figure \ref{Fig1} panels A and B - black and blue boxes). 
    \item ATLAS systematically detects objects with higher orbital inclinations than CSS (Figure \ref{Fig1} panel C - green box).
    \item ATLAS is better at detecting objects brighter than H = 19 than G96 (Figure \ref{Fig1} panel D - yellow box).
\end{itemize}

It is clear from Figure \ref{Fig1} and the previous itemized summary that ATLAS is complementary to G96. By analyzing and debiasing ATLAS telescope detections, we can provide information in the NEO population that was missed by G96, and thereby potentially improve NEOMOD2. This can be done especially in regards to the highly inclined NEOs as well as to the number (completeness) of objects brighter than H = 19, or more specifically H $<$ 17.75. The latter value has been used as a reference point for the number of km-sized NEOs.  It assumes that NEOs have an average geometric albedo of $\langle p_V\rangle=$ 0.14 \citep[see also \citet{Morbidelli2020} for small deviations]{Stuart2004}.

\begin{figure}[h!]
    \centering
    \includegraphics[width=1.\linewidth]{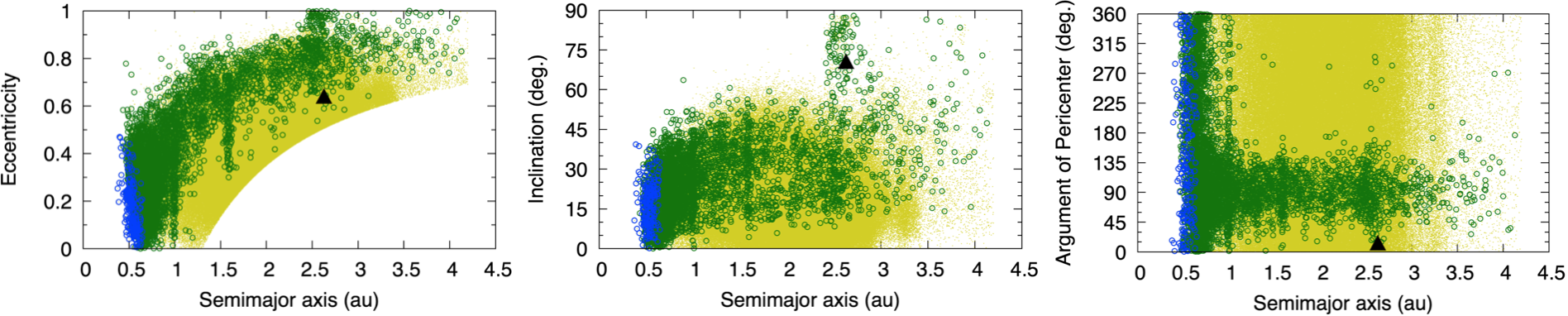}
    \caption{Eccentricity, Inclination, and Argument of Pericenter (from left to right) as a function of Semimajor axis of 350,095 NEOs with absolute magnitudes between 15 $<$ H $<$ 25 from NEOMOD2 simulator (yellow dots). Green symbols represent NEOs that were missed by all FoVs from CSS G96 over the survey observation period during years 2005 and 2023. Blue symbols are for NEOs missed by all four ATLAS telescopes (those overlap with green symbols in the same region) between 2015 and 2023. The large black triangle represents the orbital elements of NEO 2022 RX3 with $a=$ 2.626 au, $e=$ 0.64, $i=$ 70.6$\degr$, $\mathrm{w}=$ 13.1$\degr$, and H = 17.7 that was discovered by the ATLAS telescope in Chile (W68).}
    \label{Fig2}
\end{figure}

Another advantage for ATLAS over G96 is that ATLAS has full sky coverage, which allows it to detect objects that do not appear in the G96 FoVs. To better explain this concept, we did the following experiment. First, we generated a population of 350,095 NEOs with absolute magnitudes between 15 $<$ H $<$ 25 using the NEOMOD2 simulator \citep{Nesvorny2024neomod2}. At this point, we are not interested in determining the kinds of NEOs that would necessarily be detected by ATLAS or G96. Our goal is to assess which NEOs would be found in each survey's FoV during their operation interval as described previously (see also caption of Figure \ref{Fig1}). We did this by propagating the generated NEO orbits with {\tt oIF} \citep{Naidu2017} and recording all the NEOs from NEOMOD2 that overlapped the ATLAS and G96 FoVs over the simulated survey period. The results of our experiment are shown in Figure \ref{Fig2}.

It is clear from Figure \ref{Fig2} that ATLAS telescopes cover many regions of the sky where NEOs would not be observed by CSS G96. Specifically, the distribution of NEOs that missed all G96 FoVs (green symbols in Figure \ref{Fig2}) indicate that G96 has difficulties in detecting NEOs with semimajor axes less than 1 au and high eccentricities and inclinations at pericenter angles around 90$\degr$ or below (for any semimajor axis distance). These highly eccentric and inclined orbits at low to moderate argument of pericenter angles represent orbital geometries where closest approach to Earth may happen in Earth's southern hemisphere. Therefore, observing such objects that have been potentially hiding from G96 may require consistent monitoring of the southern sky, which G96 does not do. In fact, NEO 2022 RX3, with $a=$ 2.626 au, $e=$ 0.64, $i=$ 70.6$\degr$, $\mathrm{w}=$ 13.1$\degr$, and H = 17.7 (black triangle in Figure \ref{Fig2}) was discovered by the ATLAS telescope in Chile (W68) within its first year of operation. Note that 2022 RX3 is well within the orbital element regions where NEOs were hiding from G96, especially when considering their eccentricity versus semimajor axis distribution (clustered green symbols shown in Figure \ref{Fig2} center panel).

One last advantage from ATLAS compared to CSS G96 is that, once combined, ATLAS telescopes are capable of potentially observing and detecting NEOs whose orbital period ratio equals to a multiple of Earth's orbital period. The vertical clusters of green symbols in Figure \ref{Fig2} coincide with the distances where the synodic motion of NEOs allow them to hide from CSS G96 FoVs\footnote{See related discussion in ``closing comments'' on the poster by Harris \& Chodas in their ACM presentation (\url{https://www.hou.usra.edu/meetings/acm2023/eposter/2519.pdf}), where one is a figure showing that the actual impacting trajectories come strongly concentrated in the solar and anti-solar (opposition) direction, and the second comment shows the excess of ``undiscovered'' NEOs in resonant orbits, i.e., 1 yr, 2 yr, etc. periods, as discussed in this work and shown in Figure \ref{Fig2}.}. As first noted by \citet{Tricarico2017} and later emphasized by \citet{Nesvorny2023neomod,Nesvorny2024neomod2}, these distances are associated with lows (`dips') in the detection probability of the telescopes as a function of semimajor axis. This issue can be overcome by combining the FoVs and observations from all four ATLAS telescopes. Alone, each single ATLAS telescope would suffer from the kinds of synodic effects as G96 (see Section \ref{sec:performance}, Figure \ref{Fig5}). This result shows the importance of combining survey data from all four ATLAS telescopes.

Our work is structured as follows. In Section \ref{sec:model} we summarize the modeling from \citet{Nesvorny2023neomod,Nesvorny2024neomod2} for determining bias and magnitude distribution fits, now updated to perform with ATLAS observations. In this section, we also discuss the survey's limiting magnitude cutoff (Section \ref{sec:vcut}) and trailing loss effects (Section \ref{sec:trlloss}). In Section \ref{sec:results}, we report on the debiasing power of ATLAS and present our model and H distribution results. We do this as follows. Section \ref{sec:performance} is dedicated to analyzing the debiasing power of each of the four ATLAS telescopes. Our assessment of how well our model bias represents real detections from the ATLAS telescopes are presented in Sections \ref{sec:atlas01a} (Mauna Loa; T08), \ref{sec:atlas02a} (Haleakal\=a; T05),  \ref{sec:atlas03a} (Sutherland; M22), and \ref{sec:atlas04a} (Chile; W68). Results for the overall debiased H distribution, and survey completeness of all four telescopes combined, are presented in Section \ref{sec:atlasall}. Here we also compare them with previous literature findings \citep[i.e,][]{Heinze2021,Harris2021,Nesvorny2024neomod2}. Our conclusions are presented in Section \ref{sec:conclusion}.

\section{Bias and Fitting Model} \label{sec:model}

This work follows the same methodology developed in \citet{Nesvorny2023neomod,Nesvorny2024neomod2} for constructing NEOMOD 1 and 2.  Our new model will determine Model Bias and NEO H distribution model fits for the ATLAS data. The model and methods developed in \citet{Nesvorny2023neomod,Nesvorny2024neomod2} are flexible enough to be easily updated with different telescope and observational data (e.g., such as ATLAS;  this work; or 2023+ CSS G96 upcoming data, and even data from the Vera C. Rubin Observatory\footnote{\url{https://rubinobservatory.org}}, and NEO Surveyor\footnote{\url{https://neos.arizona.edu}} once they become available in future years). 

All that the method/model needs is ($i$) a well defined set of FoV pointing directions (i.e. declination, right ascension, and angle with the North pole direction -- all with respect to the center of the FoV -- as well as the associated Modified Julian Date, MJD, of the exposure, and the size and shape of the FoV), ($ii$) the telescope's efficiency at detecting objects (NEOs, main belt asteroids, etc.), during each MJD of FoV exposures, and ($iii$) the list of unique NEO detections from that telescope. All items $i$, $ii$, and $iii$ need to be consistently derived within the same period of the survey's operation.

The first step in the method consists in determining the survey's biased model as a function of $a$, $e$, $i$, and H, i.e. $\mathcal{M}_b$($a,e,i$, H). For that, we need to take the list of FoV pointing directions as described in the previous paragraph and generate a survey database.  The vast majority (over $99\%$) of the ATLAS survey's 4 million science exposures to date consist of 30-second exposures in either the ATLAS-specific `cyan' ($c$) or `orange' ($o$) filters, observed in a ``quad'' sequence of four exposures at the same sky footprint spread over 20-30 minutes.  The $o$ filter approximates combined Sloan $r+i$ filters and is the primary filter for ATLAS observations because it is relatively insensitive to scattered light from the Moon. The $c$ filter (comparable to combined Sloan $g+r$) is used only when the Moon is below the horizon.  For this work, we selected only footprints where the survey was able to complete four observations and produce a zeropoint for each of the four exposures. 

For ATLAS we obtained a total of 358,951 FoV pointing directions for the Mauna Loa (T08) site between the years 2017 and 2022, 372,257 for the Haleakal\=a (T05) site between 2015 and 2023, 65,521 and 79,600 for Sutherland (M22) and Chile (W68) sites, respectively, both between 2022 and 2023. Then, we subdivided the NEO space in 41 bins of $a$ between 0.1 au and 4.2 au with $da=$ 0.1 au, 25 bins of $e$ between 0 and 1 with $de=$ 0.04, and 22 bins of $i$ between 0$\degr$ and 88$\degr$ with $di=$ 4$\degr$ \citep{Nesvorny2023neomod,Nesvorny2024neomod2}. Finally, we populated each of those bins with 1000 test particles, which is a large enough number to statistically saturate our convergence tests \citep{Nesvorny2023neomod} due to ATLAS very large FoV dimensions \citep[5.4$\degr \times$ 5.4$\degr$;][]{Tonry_2018}.

Once our NEO space has been binned and the FoV database has been created, we use those components as input for {\tt oIF} \citep{Naidu2017}. {\tt oIF} simulates the survey we want to determine and calculate $\mathcal{M}_b$($a,e,i$, H) and records all particles that overlap with the survey's FoVs (as in our example from Section \ref{sec:intro}). The particles in overlapping FoVs have, among several associated quantities, their distance (i.e., particle-Earth, particle-Sun), their phase angle, as well as their angular velocity (apparent motion) with respect to Earth. All have this information recorded in an ASCII file, along with the observer-Sun distance at the time of FoV overlap. We then use this information to post process {\tt oIF}'s outputs by adding to each particle different values of absolute magnitude (H) in order to infer their visual magnitude (V) at the time they overlapped FoVs \citep[more details in][]{Nesvorny2023neomod}. To those calculated V magnitudes, we apply the detection efficiencies of the telescopes consistently related to each MJD of exposure of FoV overlap to determine the overall probability of detection of a NEO with magnitude H in a single bin ($a,e,i$). This is done following Eq.~(3) in \citet[derived from \citet{Jedicke2016}]{Nesvorny2023neomod}:
\begin{equation}
\epsilon({\rm V}) = \frac{\epsilon_{\rm 0}}{1+\exp\left({\frac{\rm V - V_{lim}}{\rm V_{width}}}\right)}
\label{eq1}
\end{equation}
\noindent where $\epsilon_{\rm 0}$, ${\rm V_{\rm lim}}$, and ${\rm V_{\rm width}}$ are the nightly probability of detection for point source objects, the (limiting) visual magnitude where the
probability of detection drops to 50\%, and the width of the transition to zero detection probability. All these parameters vary as influenced by night conditions \citep{Nesvorny2023neomod,Nesvorny2024neomod2} and were derived directly from ATLAS telescopes. 

To compute the nightly efficiency parameters in Eq.~(\ref{eq1}), the ATLAS survey maintains a catalog of all numbered and multi-opposition asteroids from the Minor Planet Center. The orbits of these asteroids are known to very high accuracy; their sky-plane coordinates can be predicted to within arcseconds of their actual position. ATLAS computes the position of every known asteroid for each exposure and produces a known-asteroid file for each exposure that contains all asteroids predicted to fall within the exposure and at least as bright as V $=$ 20.5, well fainter that the survey's V $\approx$ 19.7 sensitivity limit \citep{Tonry_2018}. Also stored with the asteroids predicted to be in the field are their predicted apparent magnitudes, sky-plane velocities, and their orbital elements.  

\begin{figure}[h!]
    \centering
    \includegraphics[width=1.\linewidth]{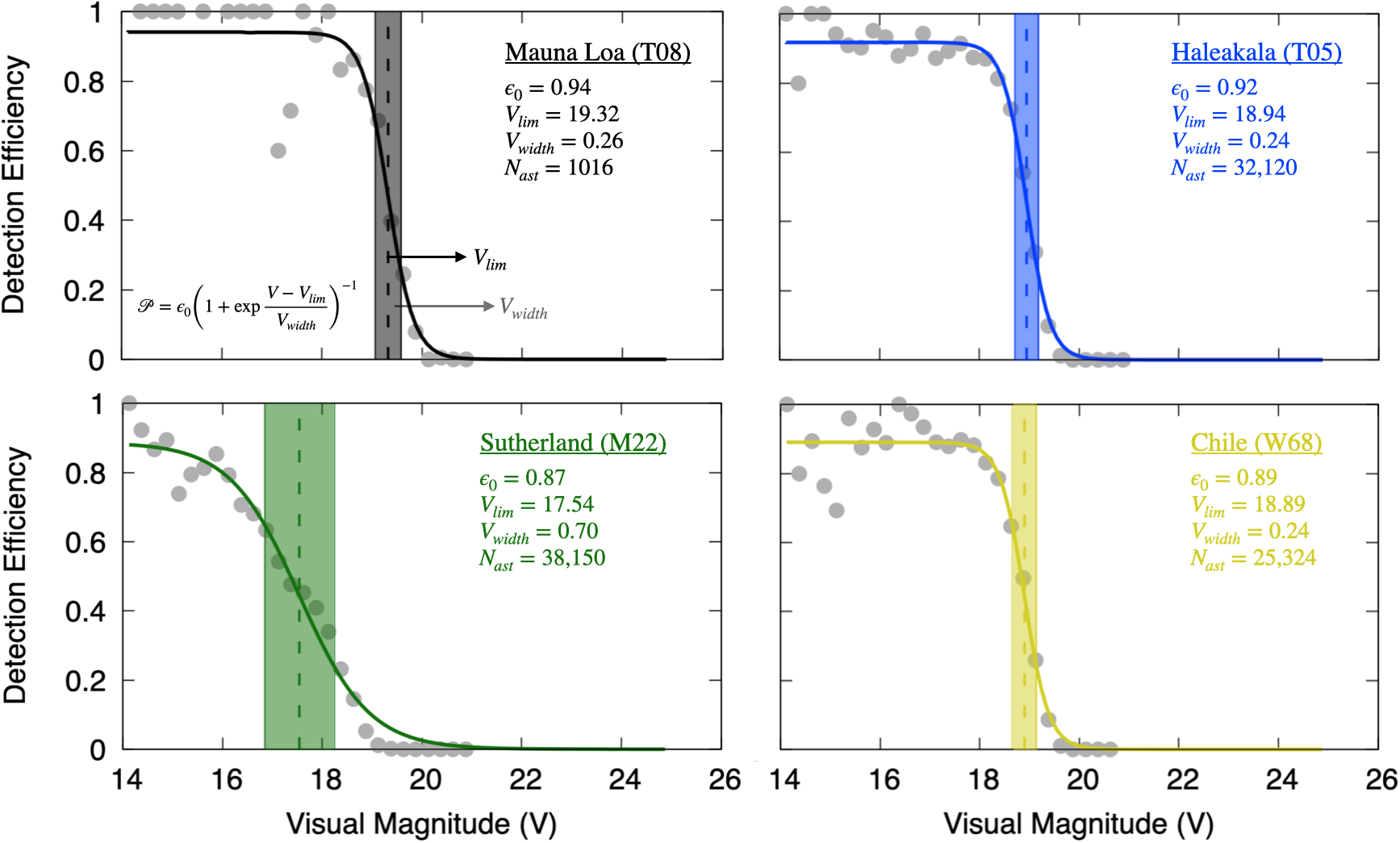}
    \caption{Detection efficiency as a function of V (Eq.~\ref{eq1}) for all ATLAS telescopes. Colors are: black for Mauna Loa (T08 -- top left), blue for Haleakal\=a (T05 -- top right), green for Sutherland (M22 -- bottom left), and yellow for Chile (W68 -- bottom right). This color coding will be kept the same throughout the paper. The vertical dashed lines represent $V_{\rm lim}$ whereas the shaded regions stand for the witdh in which the efficiency (i.e., probability) function transition from $\epsilon_{\rm 0}$ to zero. All panels were obtained for the same night. The fitted values of $\epsilon_{\rm 0}$, $V_{\rm lim}$, and $V_{\rm width}$ are reported in each panel for reference, along with the number of asteroids ($N_{ast}$) satisfying our q4 criteria on that night.}
    \label{FigEffNew}
\end{figure}

A process that executes each morning after an ATLAS telescope has completed nightly observations matches predicted asteroid positions with transient detection positions from the ATLAS image subtraction catalogs that are used to detect asteroids. If a predicted position matches a transient detection to within $\pm 3$ pixels, the asteroid is considered ``detected'' in that particular exposure. Similarly, if an asteroid is matched in all four exposures in a quad processed by the ATLAS MOPS asteroid pipeline, it is considered ``detected'' by MOPS \citep{Denneau2013}. We use  this nightly catalog of known asteroid positions to determine whether detections  were made by ATLAS. Then, we create 0.25-magnitude wide bins and store in each bin the fraction of asteroids that were detected in that magnitude bin. Finally, from this data we fit Eq.~(\ref{eq1}) using the {\tt R} programming language's {\tt nls()} (nonlinear least-squares) function and extract the $\epsilon_{\rm 0}$, $V_{\rm lim}$ and $V_{\rm width}$ parameters as follows: {\tt 
eff0, L, w = nls(effy $\sim$ eff0 / (1 + exp((effx - L) / w)), algorithm='port')}, where {\tt nls()} is the R nonlinear least squares function, and {\tt effx} is the tabulated efficiency in V magnitude bins. ATLAS produces these parameters under a variety of detection circumstances: single exposure detections, quad detections prior to asteroid pipeline processing, and quad tracklet detections after asteroid pipeline processing. Measuring performance in these different circumstances helps in understanding losses in the ATLAS detection pipeline. For this work, we use only the ``quad tracklet'' efficiency called q4, which represents the magnitude-dependent ability of the ATLAS asteroid pipeline to produce a tracklet when there are four exposures \citep[also considered only asteroids detected over four CSS exposures when computing and fitting their efficiency parameters]{Nesvorny2024neomod2}. An example of our efficiency curve fitting q4 tracklets is shown in Figure \ref{FigEffNew} for all four ATLAS telescopes.

\begin{figure}[h!]
    \centering
    \includegraphics[width=1.\linewidth]{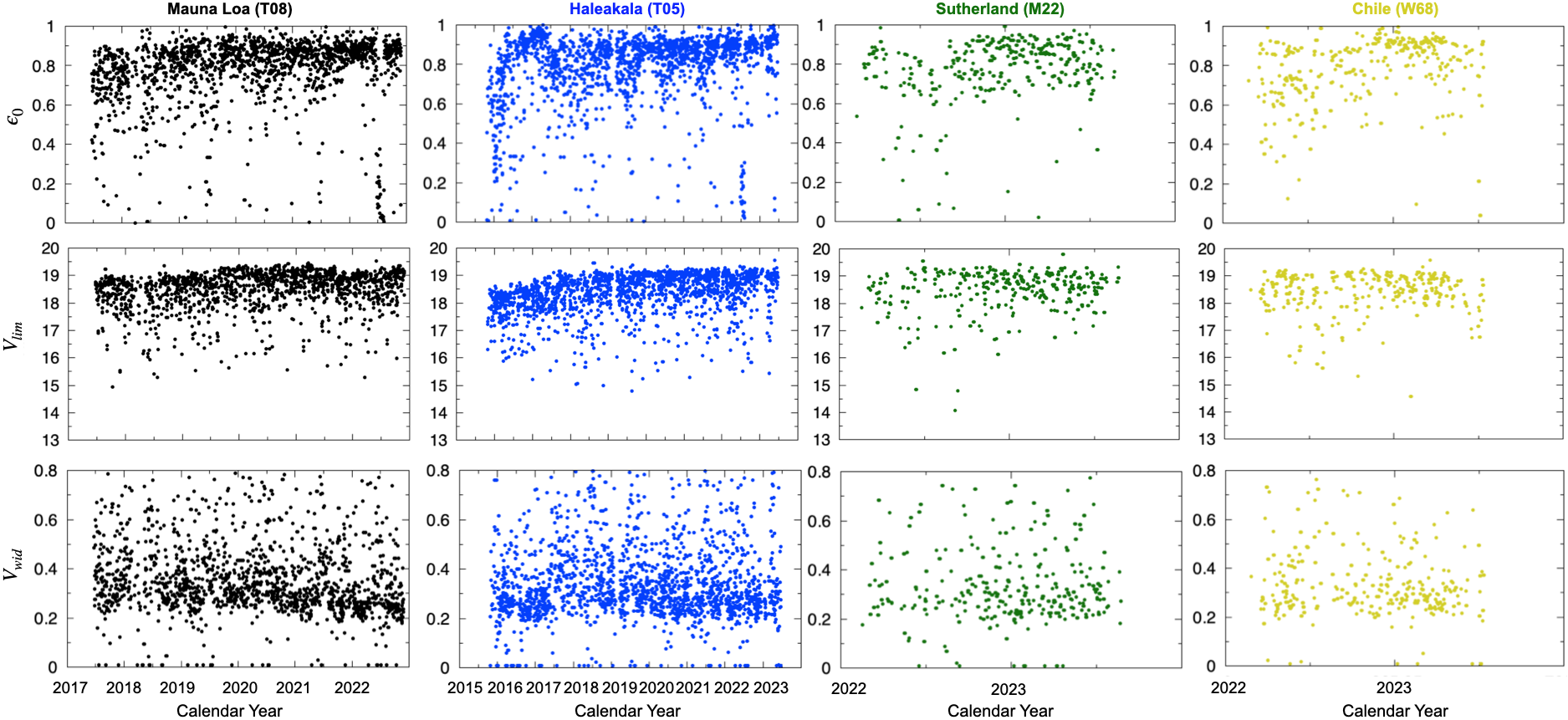}
    \caption{Variation of ATLAS efficiency parameters (Eq.~\ref{eq1}, see also Figure \ref{FigEffNew}) on a nightly basis. Colors are: black for Mauna Loa (T08), blue for Haleakal\=a (T05), green for Sutherland (M22), and yellow for Chile (W68).}
    \label{FigNew}
\end{figure}

The data shown in Figure \ref{FigEffNew} were obtained during the same night for all telescopes, therefore, differences in the fitted values reported, which affects the shape of the fitted efficiency curves for each telescope, are a result of different sky conditions at those sites. The night conditions also affect the number of observed asteroids satisfying the q4 criteria. Furthermore, on some nights, such as when a telescope is observing an area near the north or south celestial pole or when observations are curtailed due to weather, there are not enough observable asteroids to produce a fit using the {\tt nls()} routine, and therefore no statement can be made about observing efficiency. These nights have been omitted from this study. Figure \ref{FigNew} shows how $\epsilon_{\rm 0}$, $V_{\rm lim}$, and $V_{\rm width}$ vary over time (night by night) for each individual ATLAS telescope during the observational period considered in our work. 

The plots in Figure \ref{FigEffNew} are examples from individual nights at different ATLAS telescopes. In general, these fits can be quite noisy due to huge variation of numbers of detectable objects and losses from weather. For example, in the case of the top left panel for Mauna Loa (T08), the observed area was the northern sky well north of the ecliptic, and there were only 53 detectable objects with $V<18$, with some $V$ bins containing only a single object. The losses in the bins near $V=17$ pull down the $\epsilon_{\rm 0}$ fit from a plausible ``eyeball'' value near $1.0$. Similarly for the Sutherland (M22) telescope, the fitted form of Eq. \ref{eq1} deviates slightly from the binned data near the faint end at around V$_{\rm lim}$ + 2V$_{\rm width}$. This mismatch suggests there could be modest systematic losses at the very faint end of detectability for the Sutherland (M22) telescope. These differences could potentially be handled using the same methods from \citet{Nesvorny2024neomod2} in NEOMOD2, which derived a 6-parameter fitting curve based on raw observation data from G96 alone. For this work, we are analysing four telescopes at once, instead of only one as in \citet{Nesvorny2024neomod2}, and we elected to use the fitted data based on Eq. \ref{eq1} that is computed each night for all ATLAS telescopes even though the fits are noisy and may exhibit small systematic effects (Figures \ref{FigEffNew} and \ref{FigNew}). Further understanding of these system effects requires additional investigation from the ATLAS team.

Our bias determination is directly influenced by the evaluation of Eq.~(\ref{eq1}) on a nightly basis, i.e., for every FoV exposure. Besides the night conditions (already accounted for by $\epsilon_{\rm 0}$, ${\rm V_{\rm lim}}$, and ${\rm V_{\rm width}}$) that influence the determination of the overall detection probability $\mathcal{P}$($a,e,i$, H) (i.e., our determination of the biased model, which we will define below), there are two other major complications. These are related to the computation of (i) the survey's cutting visual magnitude from where we consider $\epsilon_{\rm 0}$ = 0 if V $>$ V$_{\rm cutoff}$, and (ii) the computation of trailing loss effects, e.g., smearing of tracklets (changes in trailing intensity) caused by fast moving objects, which cause them to become fainter and harder to detect. Both effects will be discussed in more detail in Sections \ref{sec:vcut} and \ref{sec:trlloss}, but they are accounted for when building our biased model ($\mathcal{M}_b$). 

Ultimately, after running over the entire range of bins ($a,e,i$) and over a large range of H, and computing $\epsilon({\rm V})$ for every object that overlapped with our FoVs, while also accounting for the effects described in Sections \ref{sec:vcut} and \ref{sec:trlloss}, we are capable of generating the survey's mean detection probability $\mathcal{P}$($a,e,i$, H) over the entire survey's period of performance.
With $\mathcal{P}$($a,e,i$, H) well characterized, we define the biased model as \citep{Nesvorny2023neomod,Nesvorny2024neomod2}

\newpage
\begin{equation}
\mathcal{M}_b(a,e,i,{\rm H}) = \mathcal{P}(a,e,i,{\rm H})\, n({\rm H})\sum_{j=1}^{n_s} \alpha_j({\rm H})\,p_{q^*,j}(a,e,i,{\rm H})
\label{eq2}
\end{equation}

\begin{itemize}
    \item $n$(H) = dN/dH is the differential absolute magnitude distribution, where its counterpart cumulative distribution will be denoted as N(H) and fitted by cubic splines to represent $\log_{\rm 10}$N(H) \citep[see][for details on splines]{Nesvorny2023neomod};
    \item $\sum_{j=1}^{n_s} \alpha_j({\rm H})=$ 1, with $\alpha_j({\rm H})$ the magnitude-dependent weights of each individual source and $n_s$ the number of model sources. We use $n_s=$ 12, being eight individual resonances ($\nu_{\rm 6}$, 3:1, 5:2, 7:3, 8:3, 9:4, 11:5, and 2:1), inner main belt weak resonances, Hungarias and Phocaeas (representing high-inclination sources), as well as Jupiter-family comets. The magnitude-dependent weights of different $n_s$ are assumed to have a linear dependency on H for simplicity, thus $\alpha_j({\rm H})=\alpha_j^{(0)}+\alpha_j^{(1)}({\rm H}-{\rm H}_\alpha)$ (H$_\alpha$ being some reference value); 
    \item $p_{q^*,j}(a,e,i,{\rm H})$ is the orbital distribution PDF of NEOs from source $j$, which also includes effects from disruption at low perihelion distance $q$ \citep{Granvik2016,Nesvorny2023neomod,Nesvorny2024neomod2}. This means that $p_{q^*,j}(a,e,i,{\rm H})$ accounts for NEOs being eliminated once reaching $q^*$(H) $\approx q^*_{\rm 0}+\delta q^*$(H-H$_q$), with H$_q=$ 20 magnitude. The normalized PDF can then be expressed, for any H, as $\int p_{q^*,j}(a,e,i,{\rm H})\,{\rm d}a\,{\rm d}e\,{\rm d}i=$ 1; 
\end{itemize}
With the model bias now quantified, we can use {\tt Multinest}\footnote{\url{https://github.com/farhanferoz/MultiNest}} \citep{Feroz2008,Feroz2009} to perform model selection via parameter estimation and error analysis.
The {\tt MultiNest} results are expressed in terms of log-likelihoods
\begin{equation}
    \mathcal{L} = -\sum_j \lambda_j + \sum_j n_j \ln\lambda_j,
\label{eq3}
\end{equation}
\noindent where $\lambda_j$ represents the number of objects in bin $j$ expected from the biased model, with the sum executed over all $a, e, i$ and H bins \citep{Nesvorny2023neomod}. The value $n_j$ is the number of unique ATLAS detections in bin $j$. There is a total of 30 model parameters: 22 coefficients defining $\alpha_j$, 6 parameters that define the magnitude distribution from splines (i.e., five slopes and the overall normalization), and 2 parameters for the size-dependent disruption ($q_0^*$ and $\delta q^*$).

Equation (\ref{eq2}) is constructed following the methods previously described in this section for each {\tt MultiNest} trial. This allows us to define $\lambda_j=\mathcal{M}_b(a,e,i,{\rm H})$ (the expected number of events) in every bin of the model domain. After millions of trials, {\tt MultiNest} converges to the maximum log-likelihood model parameters that can be used to determine the intrinsic (debiased) model \citep{Nesvorny2023neomod,Nesvorny2024neomod2}
\begin{equation}
\mathcal{M}(a,e,i,{\rm H}) = n({\rm H})\sum_{j=1}^{n_s} \alpha_j({\rm H})\,p_{q^*,j}(a,e,i,{\rm H}).
\label{eq4}
\end{equation}

\subsection{Limiting visual magnitude cutoff} \label{sec:vcut}

\begin{figure}[h!]
    \centering
    \includegraphics[width=0.8\linewidth]{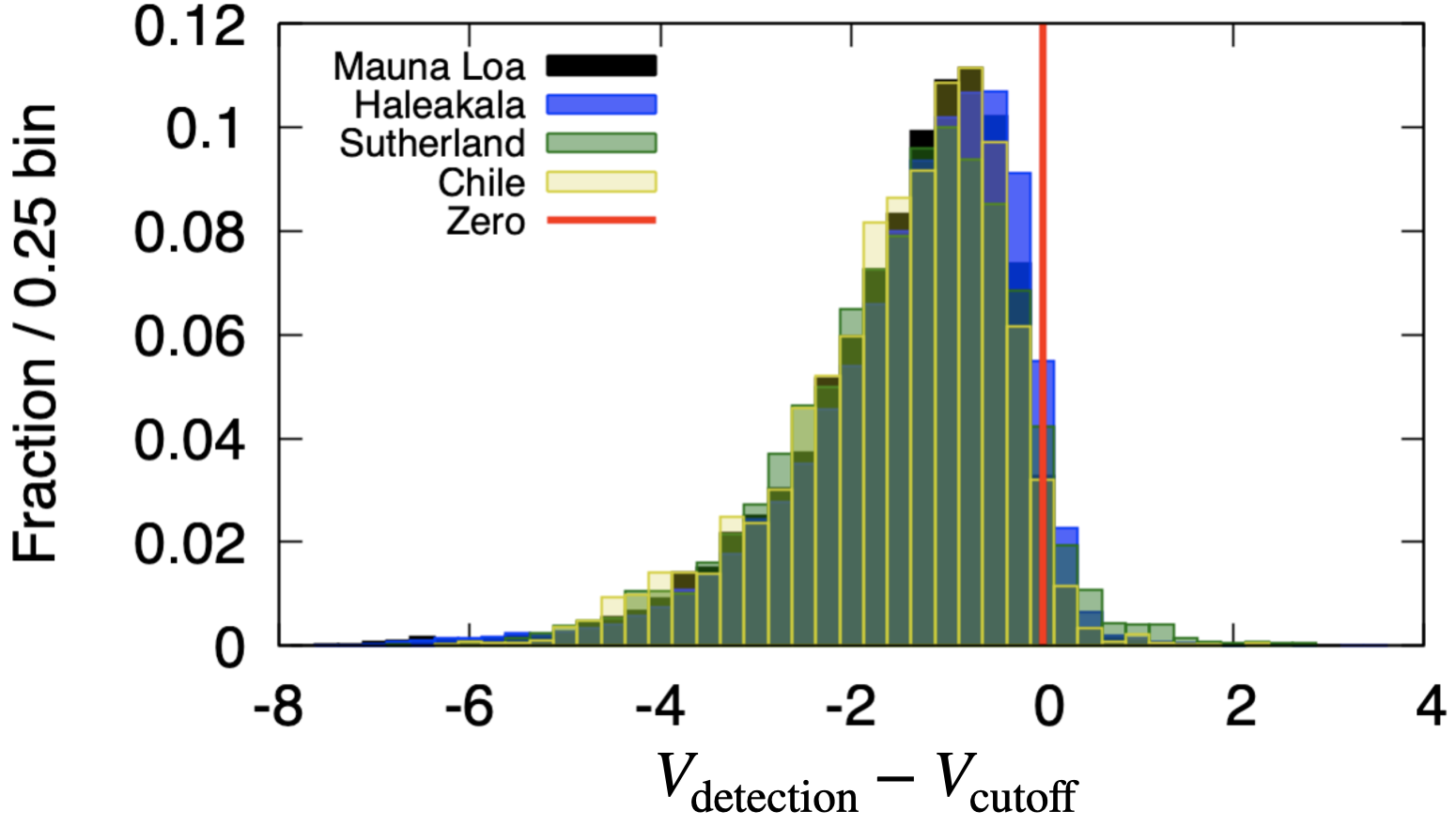}
    \caption{Fractional number of real detections by ATLAS as a function of the object's visual magnitude at the detection ($V_{{\rm detection}}$) subtracted from the survey's cutoff limiting magnitude ($V_{{\rm cutoff}}=V_{\rm lim}+V_{\rm width}$). The red vertical line indicates $V_{{\rm detection}}-V_{{\rm cutoff}}=$ 0. Different ATLAS telescopes are represented by different colors as shown by the labels in the top left corner.}
    \label{Fig3}
\end{figure}

    The ability of a survey to detect an object with visual magnitude (V) varies following the functional form presented in Eq.~(\ref{eq1}) (see also Figure \ref{FigEffNew}). The survey values of $\epsilon_0$, $V_{\rm lim}$, and $V_{\rm width}$ change on nightly basis as night conditions are not always the same (Figure \ref{FigNew}). Figure \ref{Fig3} shows the distribution of NEOs' visual magnitude when detected by ATLAS ($V_{{\rm detection}}$) with respect to what we define as a limiting cutoff in ATLAS visual magnitude detection capabilities due to night conditions ($V_{{\rm cutoff}}=V_{\rm lim}+V_{\rm width}$). The spread in the histogram bins shown in Figure \ref{Fig3} results not only from different $V_{{\rm detection}}$ values but mostly due to variations on $V_{{\rm cutoff}}$ as $V_{\rm lim}$ and $V_{\rm width}$ varies from night to night (as presented in Figure \ref{FigNew} second and third rows from the top), i.e., two objects with the same value of $V_{{\rm detection}}$ may fall in a different histogram bin if observed in different nights. 

Looking more closely at all four panels in Figure \ref{FigEffNew}, we can see that the detection probability function as defined in Eq.~(\ref{eq1}) predicts objects overlapping FoVs with V $>$ V$_{\rm lim}$+V$_{\rm width}$ (i.e., on the right of the shaded regions) may yet have some non-negligible chance (up to about 20\%) of being detected by the survey. However, when combining data from real detections made by ATLAS with the survey's efficiency parameters, which are fit every night (Figure \ref{FigNew}), Figure \ref{Fig3} shows a clear drop in the fractional number of objects detected under the condition V $>$ V$_{\rm lim}$+V$_{\rm width}$. The inability of a survey to detect objects beyond the limiting visual magnitude V$_{\rm lim}$+V$_{\rm width}$ was discussed in \citet{Jedicke2016} when analysing CSS data (both from G96 and 703). Here, we confirm that all ATLAS telescopes present similar characteristics to CSS in this regards. Following \citet{Jedicke2016}, we define V$_{\rm cutoff}=$ V$_{\rm lim}$+V$_{\rm width}$ as the limiting visual magnitude for potential detection, i.e. we set $\epsilon_0=$ 0, implying $\mathcal{P}=$ 0, if V $>$ V$_{\rm cutoff}=$ V$_{\rm lim}$+V$_{\rm width}$ \cite[][also used this approach once developing NEOMOD1]{Nesvorny2023neomod}. 

ATLAS performs observations in several different filters \citep{Tonry_2018}. In this work, we focus on observations (related FoVs) made using the `cyan' ($c$) and `orange' ($o$) filters (Section \ref{sec:model}). In these configurations, ATLAS nominal V$_{\rm lim}^{{\rm nominal}}$ is 19.7 \citep{Tonry_2018}. Yet, the distribution of V$_{\rm lim}$+V$_{\rm width}$ when analysed in similar way as in Figure \ref{Fig3} (see Figure \ref{FigNew} center row) shows a clear peak of the distribution at 19.2.
The correct choice of V$_{\rm cutoff}$ plays an important role in the definition of our model bias $\mathcal{M}_b$. If we were to use ATLAS's nominal V$_{\rm lim}^{{\rm nominal}}=$ 19.7 instead of V$_{\rm cutoff}=$ V$_{\rm lim}$+V$_{\rm width}$ (varying night by night) in our study, we would over-estimate the detection probability ($\mathcal{P}$) for large objects (e.g. H $<$ 19). An over-estimation of $\mathcal{P}$ for large objects would lead into an under-estimation of the population of such objects after comparing $\mathcal{P}$ with the real number of detections. Ultimately, this would result in the incorrect characterization of the intrinsic (debiased) model $\mathcal{M}(a,e,i,{\rm H})$ (Eq. \ref{eq4}).

\subsection{Trailing loss} 
\label{sec:trlloss}

When an object moves over the FoV of a telescope, if bright enough, it leaves an imprinted trail (tracklet) with intensity (TI) that is proportional to its apparent motion $\omega$ in the sky and the duration of the observation exposure time ${\rm t_{exp}}$ (i.e., TI $\propto \omega \times {\rm t_{exp}}$). The faster the object moves through the FoV, i.e. the larger $\omega$ becomes, the longer the tracklet becomes for constant ${\rm t_{exp}}$. When $\omega$ grows large enough that the object's trailed length starts crossing multiple pixels across the telescope camera, the intensity of the tracklet begins to smear out, making the object look fainter then it would be if it were resolved in a single pixel. Additionally, tracklets too large may also become impossible to be detected by their recognition techniques \citep[i.e. MOPS;][]{Denneau2013}. This effect, that can be interpreted as an increase of the objects visual magnitude V during FoV overlap, is often referred to as trailing loss.

\begin{figure}[h!]
    \centering
    \includegraphics[width=0.5\linewidth]{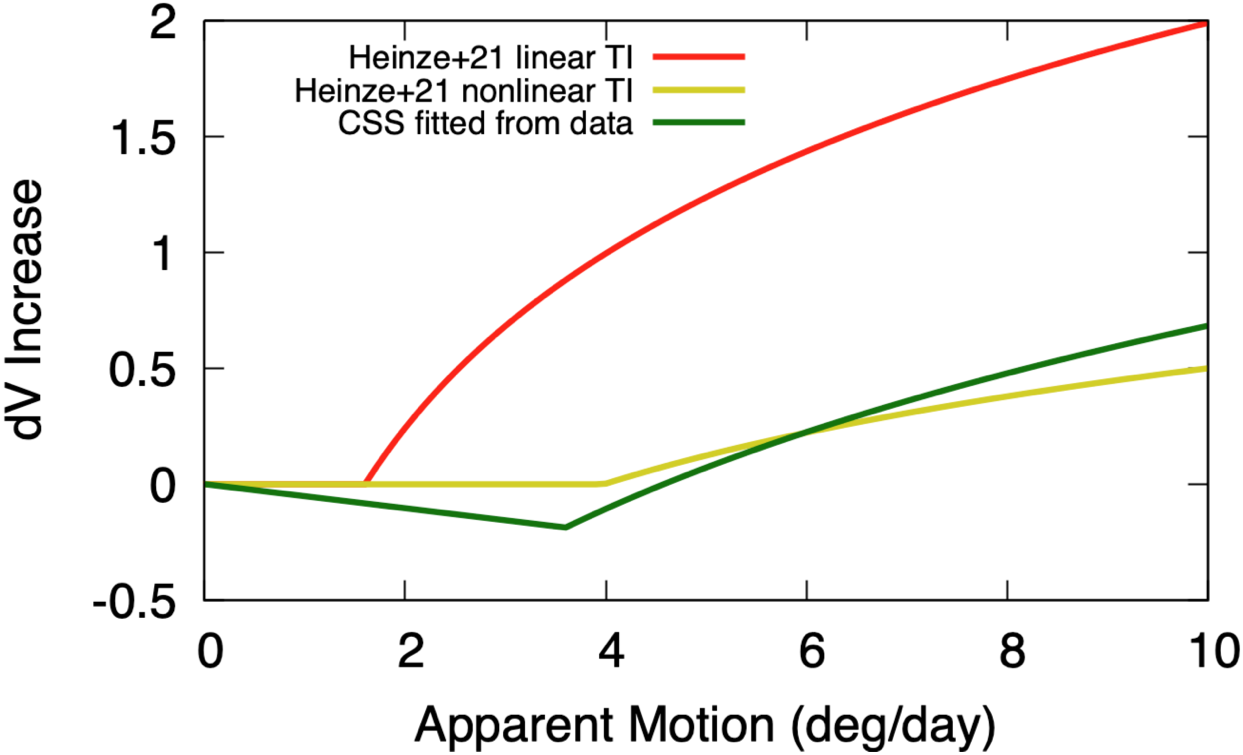}
    \caption{Trailing loss effect, i.e. increase (dV) in visual magnitude, as a function of NEOs' apparent motion ($\omega$; see main text detail in each case). Red line illustrates the effects as considered by \citet{Heinze2021}, where TI decreases linearly with $\omega$. Yellow stands for \citet{Heinze2021}'s loose estimation of ATLAS real trailing effects based on detection of objects moving with apparent motion in the range 4 $<\omega<$ 12 deg~day$^{-1}$, where TI should vary proportionally with the square root of $\omega$. Green is the trailing loss function as derived by \citet{Nesvorny2024neomod2} from real CSS data.}
    \label{Fig4}
\end{figure}

ATLAS has a pixel size and point spread function (PSF)\footnote{\url{https://atlas.fallingstar.com/specifications.php}} of about 1.86" and 3.8" respectively, and a constant FoV fixed exposure time ${\rm t_{exp}}=$ 30 seconds \citep{Tonry_2018}. Assuming that the flux of light $\phi$ that comes through the telescope camera is proportional to the trail intensity (TI) scaled by the camera pixel size, \citet{Heinze2021} suggested that $\phi_{\rm ATLAS}=$ TI/2", where 2" is the characteristic pixel size for ATLAS. The increase in visual magnitude dV that the trailed tracklet would experience is dV $\approx$ 2.5~$\log_{\rm 10}\phi$ \citep{Muinonen2010}. Therefore, an object moving at an apparent motion $\omega=$ 1.6 deg~day$^{-1}$ would leave an imprint with TI $= \omega \times$30~s $=$ 2 arcsec, which would result in a flux $\phi=$ 1 and dV = 0. This means that ATLAS telescopes should not suffer from trailing losses for apparent motions slower than 1.6 deg~day$^{-1}$. Objects moving faster than 1.6 deg~day$^{-1}$ should experience trailing loss proportional to the linear decrease of TI (i.e. as for V, larger values of TI imply in longer and fainter tracklets). For example, an object moving at $\omega=$ 16 deg~day$^{-1}$ would have TI $= \omega \times$30~s $=$ 20 arcsec, which in turn would yield a flux $\phi=$ 10 and dV = 2.5 mag.
Figure~\ref{Fig4} (red) illustrates how the increase dV in visual magnitude (trailing loss effect) varies as a function of apparent motion $\omega$ under the considerations made by \citet[see their Section 5.3]{Heinze2021}.

A linear decrease of the TI, using the apparent motion as adopted by \citet{Heinze2021}, implies an aggressive trailing loss effect. We consider this approach to be overly conservative for representing the survey's ability of detecting fast moving objects. Indeed, \citet{Heinze2021} reported that analysis for real ATLAS detections indicates that the survey's effective ability for detecting fast moving objects is considerably less aggressive. They estimate (but did not fully quantify) that the decrease in TI should be much slower, in a non-linear fashion, and proportional to the square root of the apparent motion of the NEO being observed in the range 4 $\lesssim \omega \lesssim$ 12 deg~day$^{-1}$, with dV $\approx$ 0.5 at $\omega=$ 10 deg~day$^{-1}$. This means that the effective flux $\phi_{{\rm eff}}$ can be translate as $\phi_{{\rm eff}} \propto$ TI/$\sqrt{\omega}=\kappa$~TI/$\sqrt{\omega}$, with $\kappa \approx$ 1443.42 deg$^{-1}$day$^{-1}$ a normalization constant to correct for units when considering TI in units of degree and $\omega$ in deg~day$^{-1}$. For ${\rm t_{exp}}=$ 30 seconds we then have dV $\approx$ 2.5~$\log_{\rm 10} {\rm (0.501}\sqrt{\omega}$), which is represented by the yellow curve in Figure \ref{Fig4}, where dV = 0 for $\omega\lesssim$ 3.984 deg~day$^{-1}$. This value of $\omega$ is in good agreement with the predicted 4 deg~day$^{-1}$ value reported by \citet{Heinze2021} from where ATLAS should effectively start suffering from trailing loss effects.

The less aggressive estimate of the effective trailing loss made by \citet{Heinze2021} is in close agreement with the trailing loss function derived by \citet[green in Figure \ref{Fig4}]{Nesvorny2024neomod2}, who were able to quantitatively fit trailing effects directly from the Mt. Lemmon G96 telescope data (and after the G96 camera update) over one decade of observations. 

\citet{Nesvorny2024neomod2} found that, despite the fact that fast moving objects tend to increase dV, if a NEO is not moving fast enough, it may be confused with faint stars.  This means the object may be difficult to recover by tracklet identification pipelines.  
There is, therefore, a range of apparent motion values where the object's motion in the sky would likely facilitate detection. Following \citet{Nesvorny2024neomod2}, this occurs for apparent motions between 0 $< \omega < \omega^*$ and between $\omega^* < \omega < \omega_0$. Here,  $\omega^*$ stands for the apparent motion threshold from which dV changes from decreasing to increasing, and $\omega_0$ for when dV = 0, i.e., when dV transition from negative to positive increments. Within the aforementioned range of $\omega$, the resultant effect can be understood as a decrease in the visual magnitude V of the NEO during FoV overlap (i.e. dV $<$ 0). This paradoxical effect goes against the common assumption that dV = 0 for $\omega < \omega^*$ or $\omega_0$. The optimal condition for tracklet identification in fact occurs at $\omega^*$, where dV transition from decreasing to increasing.

The Mt. Lemmon (G96) telescope camera after being updated has a pixel size and PSF of about 1.52" and 3", respectively \citep{Nesvorny2024neomod2}. These values are similar to those of ATLAS \citep{Tonry_2018}. Therefore, in this work, given the similarities between the new camera of G96 and the camera characteristics of ATLAS, as well as similarities between the yellow and green curves in Figure \ref{Fig4}, we opt for using the functional form derived in \cite{Nesvorny2024neomod2} to account for trailing loss effects. Remember that this functional form was not estimated, but instead was calibrated and fit from real data.
Such functional form \citep[see Eqs.~(3) and (4) in][]{Nesvorny2024neomod2} can be directly translated from V$_{\rm lim}$ to dV as:
\begin{align}
dV &= -A\,\omega  &{\rm for} ~\omega < \omega^*~\nonumber\\
dV &= -A\,\omega^* + 2.5\log_{\rm 10}[1+C(\omega-\omega^*)] &{\rm for}~ \omega > \omega^*,\label{eq5}
\end{align}
\noindent with the unit normalization constants A = 0.052 day~deg$^{-1}$ and C = 0.192 day~deg$^{-1}$. The transition where trailing effects start to increase dV was found to be $\omega^*=$ 3.6 deg~day$^{-1}$. Those values lead dV $>$ 0 occurring at $\omega > \omega_0 \approx$ 4.580 deg~day$^{-1}$ and dV($\omega=$ 10 deg~day$^{-1}$) $\approx$ 0.68, which values are also in close agreement with those loosely suggested by \citet{Heinze2021}, i.e. $\omega^*=\omega_0=$ 4 deg~day$^{-1}$ and dV($\omega=$ 10 deg~day$^{-1}$) = 0.5.

Trailing losses have a larger effect over faint NEOs that can only be detected once they have close approaches with Earth (i.e., they have large apparent motions across the sky). By assuming an aggressive trailing loss, we would largely penalise the detection probability $\mathcal{P}$ of the faint NEOs. That would result in an over-estimation of that population once comparing $\mathcal{P}$ with real observations. Ultimately, this factor would feed into the characterization of the intrinsic (debiased) model $\mathcal{M}(a,i,e,{\rm H})$ (Eq.~\ref{eq4}). By adopting a less aggressive trailing loss function, we expect our results to estimate a lower number of faint and small NEOs than that reported by \cite{Heinze2021}. 

\section{Results} \label{sec:results}

In our approach, each telescope can be treated as a different survey. This provides an opportunity to work toward the final goal of determining the overall bias-corrected population of NEOs from ATLAS telescopes in several steps. First, we analyze how each of the ATLAS sites performs individually (Section \ref{sec:performance}), giving a close look on how well each of the individual biased models reproduces detections. This is reported in Sections.~\ref{sec:atlas01a}, \ref{sec:atlas02a}, \ref{sec:atlas03a}, and \ref{sec:atlas04a}. These partial results allow us to understand strengths and weaknesses of each site, and also spot potential troubles in datasets (if existed). After making sure the analysis of the individual telescopes make sense, we then proceed towards combining all of them into a final bias-corrected NEO model from the ATLAS survey (Section \ref{sec:atlasall}).

\subsection{ATLAS debiasing power}\label{sec:performance}

\begin{figure}[h!]
    \centering
    \includegraphics[width=0.8\linewidth]{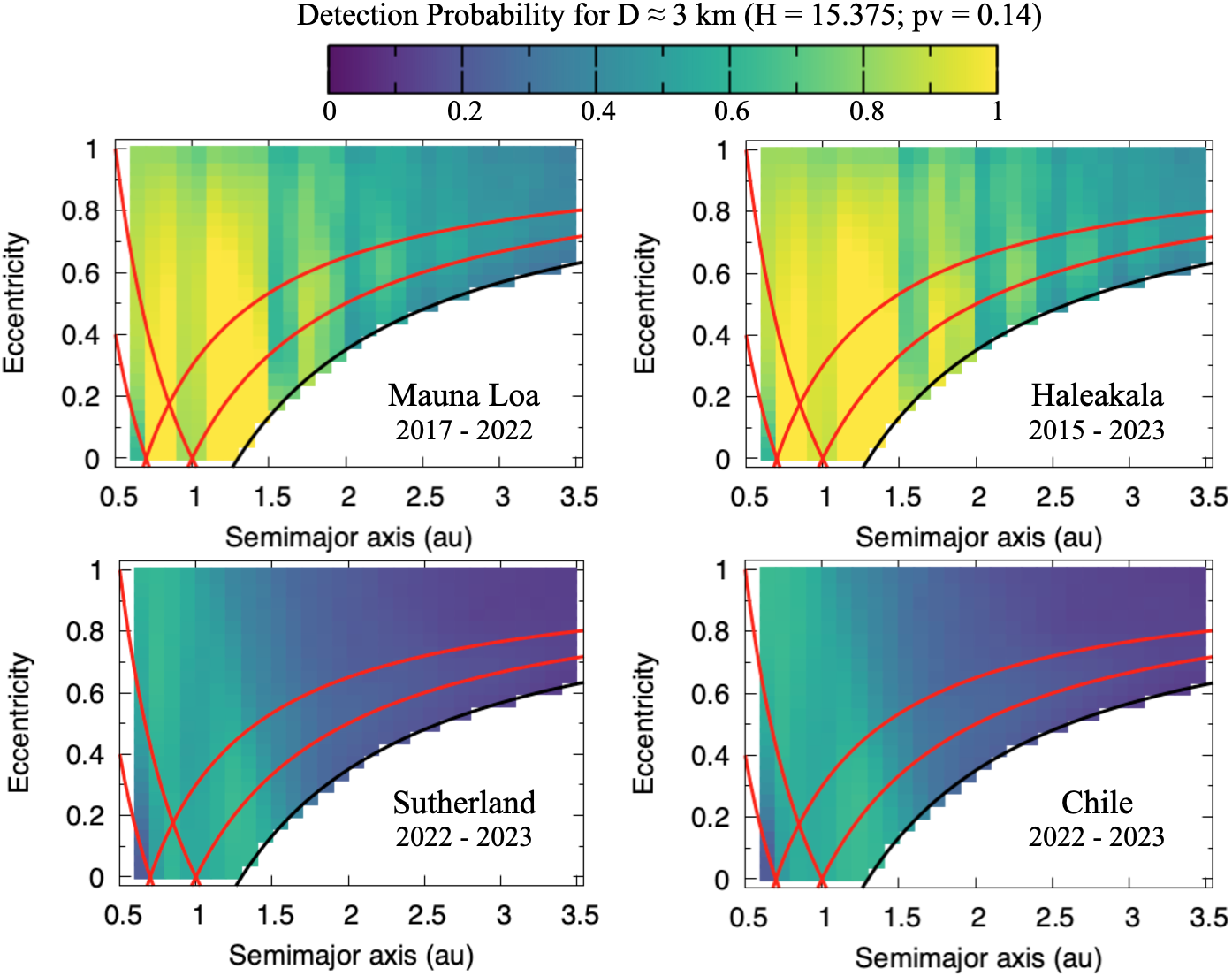}
    \caption{Averaged detection probability (color bar on top) mapped over eccentricity as a function of semimajor axis for our binned $(a,e,i)$ NEO space (Section \ref{sec:model}). The maps were generated assuming a representative bright NEO with H = 15.375, which correspond to D $\approx$ 3 km for a geometric albedo $p_V=$ 0.14. All panels are normalized over the same interval represented by the color bar. Each panel is labeled by the telescope site and years of survey operation (survey length) considered in our study. Red lines represent Venus and Earth crossing orbits. Black line stands for $q=$ 1.3 au, the limit of the NEO space.}
    \label{Fig5}
\end{figure}

To evaluate the debiasing power of each of the four ATLAS telescopes individually, we first analyze their expected average detection probabilities as a function of eccentricity and semimajor axis (Figure \ref{Fig5}). This is done by averaging our model bias $\mathcal{M}_b(a,e,i,{\rm H})$ over all inclination bins for a specific value of H, and then projecting the results on an eccentricity versus semimajor axis plane.

Figure \ref{Fig5} shows such results once considering H = 15.375, which roughly represents an NEO with diameter D $\approx$ 3 km when assuming a geometric albedo $p_V=$ 0.14. The first thing that should be noted in Figure \ref{Fig5} are the vertical stripes associated with the synodic motion of NEOs. Here the NEOs are hiding from the surveys FoVs as they orbit the Sun in multiples of Earth's orbital period \citep{Tricarico2017,Nesvorny2023neomod,Nesvorny2024neomod2}. As discussed in Section \ref{sec:intro} (Figure \ref{Fig2}; clustering of green dots hiding from CSS compared to blue dots) this effect is not observed when combining FoVs for all ATLAS telescopes, but they become apparent when analyzing each telescope on its own. Thus, the importance of understanding the behavior of each ATLAS telescope becomes paramount; they do not act the same, so their results must be combined in a careful manner to debias the NEO population.

\begin{figure}[h!]
    \centering
    \includegraphics[width=0.6\linewidth]{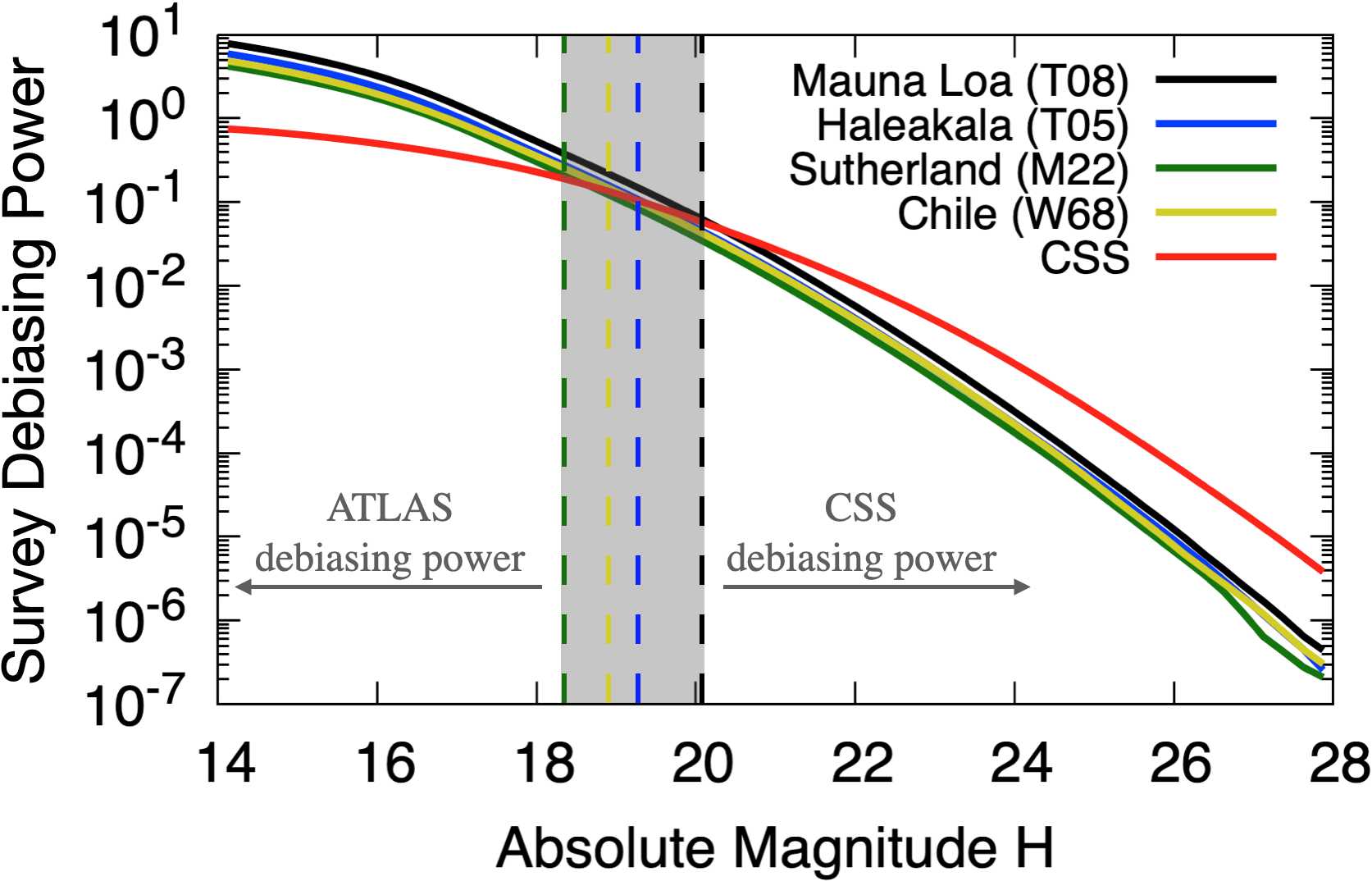}
    \caption{The power of each survey in debiasing the NEO population (see main text for details) as a function of absolute magnitude H. The individual curves report on each survey debiasing power at different values of H, measured as the averaged rate of detections (i.e., number of times an object of magnitude H is expected to be detected during the entire duration of the survey averaged over all $a,e,i$ bins) normalized by the total survey period of operation studied. In other words, the y-axis value shows how many times a single object with a given $H$ magnitude would be observed in a year of operations by a given survey telescope. Each colored line (top right labels) represents an individual survey. Vertical dashed lines indicate where each of ATLAS curves (black, blue, green, an yellow) intersects the red line for CSS. The gray shaded area delineates the range of H where all surveys have similar debiasing power. Lines {\it do not} report on which survey is better than the other at detecting objects with different H magnitudes, but rather on which may more reliably debias the NEO population at different H magnitude values according to our modeling.}
    \label{Fig6}
\end{figure}

Figure \ref{Fig5} also shows that both of the northern ATLAS telescopes, Mauna Loa (T08) and Haleakal\=a (T05), have higher averaged detection probabilities over the southern ATLAS telescopes, Sutherland (M22) and Chile (W68). We note that this does not imply that the northern telescopes are more capable, or that they are better performers, than those in the southern hemisphere (Figure \ref{Fig6}). Instead, the higher averaged detection probabilities associated with the northern telescopes are simply due to their longer period of operation (i.e., five to seven years compared to only one year), which in turn means they have had a better chance of detecting NEOs. Therefore, it is expected that the southern telescopes will become better characterized as they operate into the future.

In order to understand how well one survey performs over the other with the specific goal of debiasing the NEO population, we need to understand how efficiently they operate over similar base timelines and how reliable they are when debiasing the NEO population for different H magnitude values. Figure \ref{Fig6} reports the debiasing power of each survey for different values of H, measured as the averaged rate of detections (i.e., number of times an object of magnitude H is expected to be detected during the entire duration of the survey averaged over all $a,e,i$ bins) normalized by the total survey period of operation studied. It is important to stress, however, that what Figure \ref{Fig6} shows is how powerful a given survey can be compared to another survey in terms of debiasing the NEO population. Figure \ref{Fig6} {\it does not} show which survey is better at detecting objects with different H magnitudes, at different apparent motion speeds, or at difference values of Minimum Orbit Intersection Distance (MOID).

Figure \ref{Fig6} indicates that ATLAS telescopes have greater debiasing power at bright magnitudes (i.e. H $\lesssim$ 18--20), whereas CSS debiasing power is superior for fainter magnitudes. The 18 $\lesssim$ H $\lesssim$ 20 range reflects what is observed in Figure \ref{Fig1} (panel D - yellow box); for H = 19, the number of ATLAS detections are similar to those of CSS, for brighter H values, ATLAS overcomes CSS, and for fainter H values, CSS excels. This result is not surprising for the following reasons: CSS G96 is a much larger telescope than any of the ATLAS telescopes and can detect objects at $1.5\sim2$ magnitudes fainter than ATLAS. G96 will naturally see many more small NEOs and therefore have greater statistical debiasing power. ATLAS, however, can observe larger objects in parts of the sky that G96 cannot reach and at less favorable phase angles due to its all-sky coverage. At the zeroth order, our models for ATLAS and CSS are working reasonably.  It also implies that ATLAS is potentially a more reliable source than CSS for estimating the completeness of brighter (and larger) NEOs. 

We caution the reader that the data presented in Figure \ref{Fig6} are heavily averaged to compare all surveys to the same standards, whereas Figure \ref{Fig1} is not averaged and instead report on the unique detections made by all ATLAS and CSS telescopes over their entire survey lifetime. This means that there is nothing that prevents ATLAS from detecting more and more NEOs with H $>$ 19 in future years (recall that CSS length of operation is much longer and that the large majority of detection for H $>$ 25 was made after the Mt. Lemmon G96 camera update in May 2016). The fall-off of ATLAS curves (black, blue, green, and yellow) in Figure \ref{Fig6} compared to CSS (red) for H $>$ 20 should also be taken with a grain of salt. ATLAS is designed to detect NEOs \emph{when they are very close to the Earth}. Of the eight asteroids discovered by surveys prior to impact, ATLAS has detected three of the six that have occurred since survey inception: 2018~LA (D $\approx$ 3 m), 2019~MO (D $\approx$ 3 m) and 2022~WJ$_{\rm 1}$ (D $\approx$ 1 m). The other three were very small and undetectable by ATLAS because of their several-hour observability window prior to impact over northern Europe. Asteroid 2020~VT$_{\rm 4}$ (D $\approx$ 8 m) was discovered by ATLAS just prior to passing within 370 km of the Earth's surface. ATLAS has also demonstrated its ability to detect faint NEOs with very fast apparent motion $\omega$ and small MOID \cite[e.g., 2017 SU$_{\rm 17}$ at H = 28, $\omega\approx$ 6.43 deg~day$^{-1}$ and MOID $\approx$ 0.00061 au detected at Haleakal\=a (T05) on September 26, 2017;][see also additional examples in their Figure 1 related to the detections of 2020 BF6, 2019 TN5, and 2020 OH]{Heinze2021}. While not exhaustive, these examples demonstrate the ``Last Alert'' part of the ATLAS acronym.

Moreover, ATLAS, as an all sky survey, by design spends a lot of time looking at places on the sky where objects do not appear, and that pushes down its power for debiasing the population of small objects (H $>$ 19) as shown in Figure \ref{Fig6}. If ATLAS telescopes were to spend a greater fraction of their time on the ecliptic and opposition and/or increase its exposure time, there would be more opportunities to see small objects and ATLAS debiasing power would improve on the faint end of the H distribution. The ATLAS mission, however, is to detect close-approaching asteroids that could come from any direction, not maximize its debiasing capability. For the larger objects (H $<$ 19), the wide sky coverage helps ATLAS, because the surveys design allows for seeing these objects at different geometries and not only when they are near opposition.

Despite the northern telescopes of ATLAS having a higher averaged detection probability (i.e., better characterized) than their southern siblings, Figure \ref{Fig6} also shows that when compared to the same standards (i.e. excluding the survey period of performance dependency), all four ATLAS telescopes perform similarly. Based on this standard, it is fair to assume that over the next several years, the southern telescopes of ATLAS will improve their detection probabilities and potentially find many new NEOs. It will be interesting to revisit these issues in the future.

Having confidence that our modeling is reasonably accurate in reproducing the overall trends observed in Figures \ref{Fig1} and \ref{Fig2}, it is now important to understand how each individual ATLAS telescope operates, as well as what the debiased population expected from each of them. The following sections are dedicated to such, where we present our individual results for Mauna Loa (T08), Haleakal\=a (T05), Sutherland (M22), and Chile (W68) in Sections \ref{sec:atlas01a}, \ref{sec:atlas02a}, \ref{sec:atlas03a}, and \ref{sec:atlas04a}, respectively, while also comparing them with NEOMOD2 \citep{Nesvorny2024neomod2} findings.

\subsection{Mauna Loa (T08)} \label{sec:atlas01a}

\begin{figure}[h!]
    \centering
    \includegraphics[width=1.\linewidth]{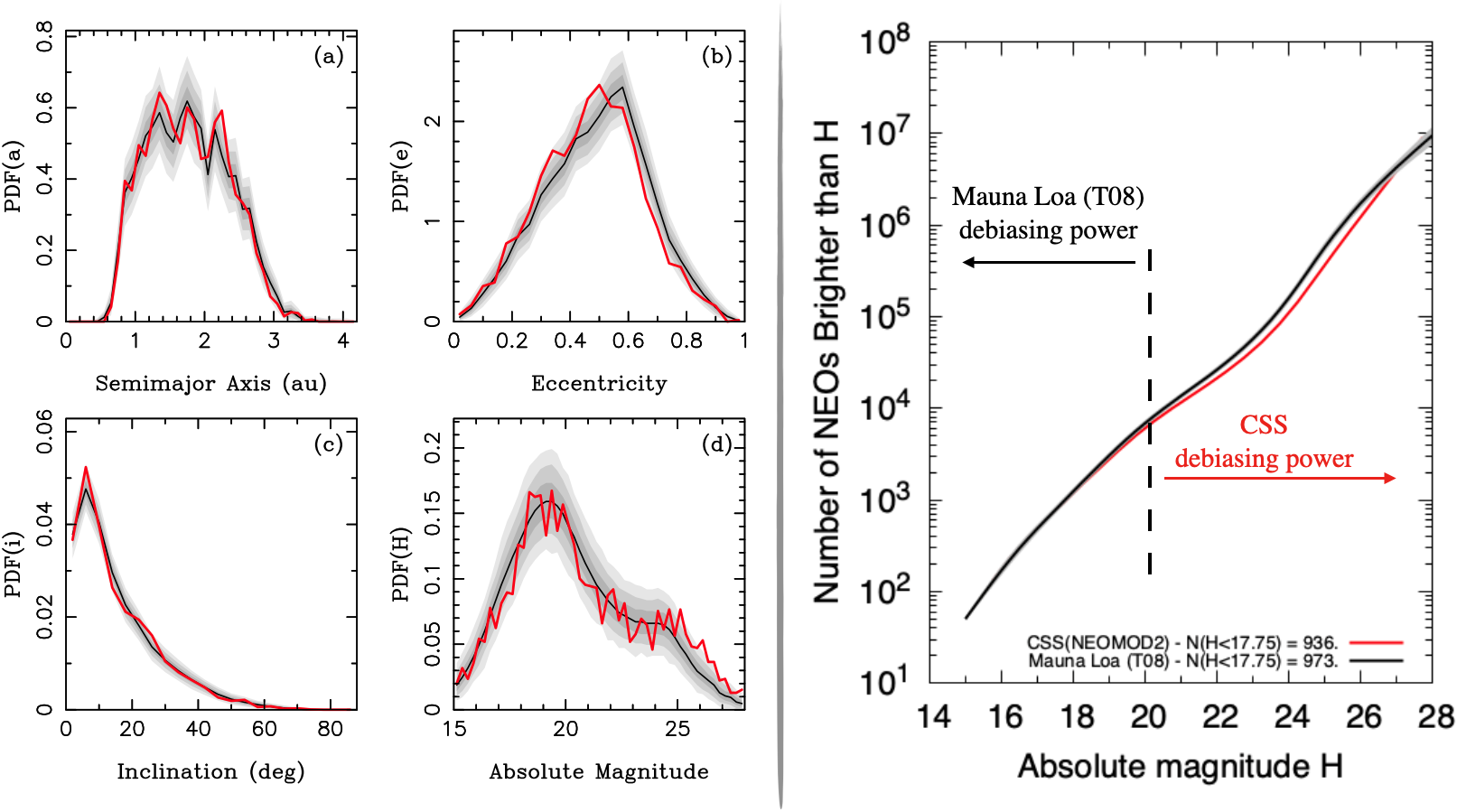}
    \caption{Four panels in the left of the gray vertical line (a, b, c, and d): The PDFs of $a,e,i$, and H from our biased base best-fit model (black lines) and Mauna Loa (T08) NEO unique detections (red lines). The shaded areas are 1$\sigma$ (bold gray), 2$\sigma$ (medium), and 3$\sigma$ (light gray) envelopes. We used the best-fit solution (i.e., the one with the maximum likelihood) from the base model and generated 30,000 random samples with 3395 NEOs each (the sample size identical to the number of Mauna Loa (T08)’s NEOs in the model domain; 15 $<$ H $<$ 28). The samples were biased and binned with the standard binning. We identified envelopes containing 68.3\% (1$\sigma$), 95.5\% (2$\sigma$) and 99.7\% (3$\sigma$) of samples and plotted them here. Right of the gray vertical line: The intrinsic (debiased) absolute magnitude distribution of NEOs from our base model (black line is the median) is compared to the magnitude distribution from \citet[red line]{Nesvorny2024neomod2}. The gray area is the 3$\sigma$ envelope obtained from the posterior distribution computed by {\tt MultiNest}. It contains 99.7\% of our base model posteriors. The vertical black dashed line indicates the approximate location of the H value where both Mauna Loa (T08) and CSS are expected to have similar debiasing power, according to what was reported in Figure \ref{Fig6} (same `survey colors' from Figure \ref{Fig6} are used in all Figures \ref{Fig7} - black, \ref{Fig8} - blue, \ref{Fig9} - green, \ref{Fig10} - yellow). Labels in the bottom of this plot report on the expected cumulative number of objects with H $<$ 17.75 (D $\approx$ 1 km for $p_V=$0.14) for both Mauna Loa (T08) and CSS (NEOMOD2).}
    \label{Fig7}
\end{figure}

The first issue to check in our modeling is how well our biased model $\mathcal{M}_b(a,e,i,{\rm H})$ is able to reproduce the survey's unique detections. Figure \ref{Fig7} shows the comparison of the Probability Density Function (PDF) for our biased model with unique detections from Mauna Loa (T08)  (four panels -- a, b, c, and d -- on the left side of the gray vertical line). Those PDFs were obtained using our {\tt MultiNest} best-fit biased-model solution (i.e., the one with the maximum likelihood) and Mauna Loa (T08)'s unique detections over the same NEO domain 15 $<$ H $<$ 28.

Mauna Loa (T08) real detection PDFs are closely reproduced by our biased model PDFs within a 3$\sigma$ envelope. Even very fine structures like the `peaks and dips' in the semimajor axis panel (a) show a good match between model and data. Those features represent how the synodic effects of NEOs avoiding FoVs \citep[Figure \ref{Fig5} top left; see][]{Tricarico2017,Nesvorny2023neomod,Nesvorny2024neomod2} affect detections at Mauna Loa (T08).

We use our best-fit biased-model $\mathcal{M}_b(a,e,i,{\rm H})$ solutions, and the 3$\sigma$ envelope (representing 99.7\% of our base model posteriors) along with the overall detection probability $\mathcal{P}(a,e,i,{\rm H})$ related to Mauna Loa (T08), to estimate its intrinsic (debiased) population model $\mathcal{M}(a,e,i,{\rm H})$ (Section \ref{sec:model}). From this debiased population we then generate the expected cumulative distribution of NEOs brighter than H (Figure \ref{Fig7} -- rightmost panel).

The number of NEOs brighter than H from our Mauna Loa (T08) debiased population is very similar to the one estimated from NEOMOD2, especially for H $\lesssim$ 20, where the debiasing power of the NEO population by Mauna Loa (T08) is higher. For reference, we estimate that the number of NEOs brighter than H = 17.75 (D $\approx$ 1 km for $p_V=$0.14) according to Mauna Loa (T08)'s debiased population is about 973$^{+47}_{-43}$. When compared to the number of NEOs currently listed in the MCP database as of February 3, 2023\footnote{Our photometric efficiency parameters, $\epsilon_0$, $V_{\rm lim}$, and $V_{\rm width}$ from Section \ref{sec:model} were derived using H magnitudes from the MPC Catalog as of February 3, 2023. Due to the fact that the absolute magnitude of NEOs listed in the MPC may change over time due to updates in their computation pipeline \citep[e.g.,][]{Pravec2012}, we can only estimate completeness in our work based on the specific catalog that we used for deriving our photometric efficiency parameters. For this reason, when comparing our estimated completeness with previous works, we use those works' reported completeness. We do not attempt to recompute them.} (850 objects), this reflects a population completeness level (N$_{\rm MPC}$/N$_{\rm T08}$) of $\approx$ 87\%$^{+4\%}_{-4\%}$. This implies that, according to Mauna Loa (T08)'s observations, within uncertainties, the NEO population to this H limit is roughly as complete as suggested by CSS \citep[i.e., 91\%$^{+4\%}_{-4\%}$ see Table 3 in][]{Nesvorny2024neomod2}. 

We do not make any detailed comparisons between Mauna Loa (T08) and CSS for H $>$ 20 because, as shown in Figure \ref{Fig6} and indicated in Figure \ref{Fig7}, CSS predictions of the debiased NEO population in this absolute magnitude range are likely to be more reliable. With that said, it is striking how close the two cumulative distributions are to one other. This is especially true once comparing each survey's number of unique detections in the range 20 $<$ H $<$ 28, i.e. 8148 for CSS (only considering detections after G96 camera update) compared to 1737 done by Mauna Loa (T08). This reaffirms the general robustness of our method.

\subsection{Haleakal\=a (T05)} \label{sec:atlas02a}

\begin{figure}[h!]
    \centering
    \includegraphics[width=1.\linewidth]{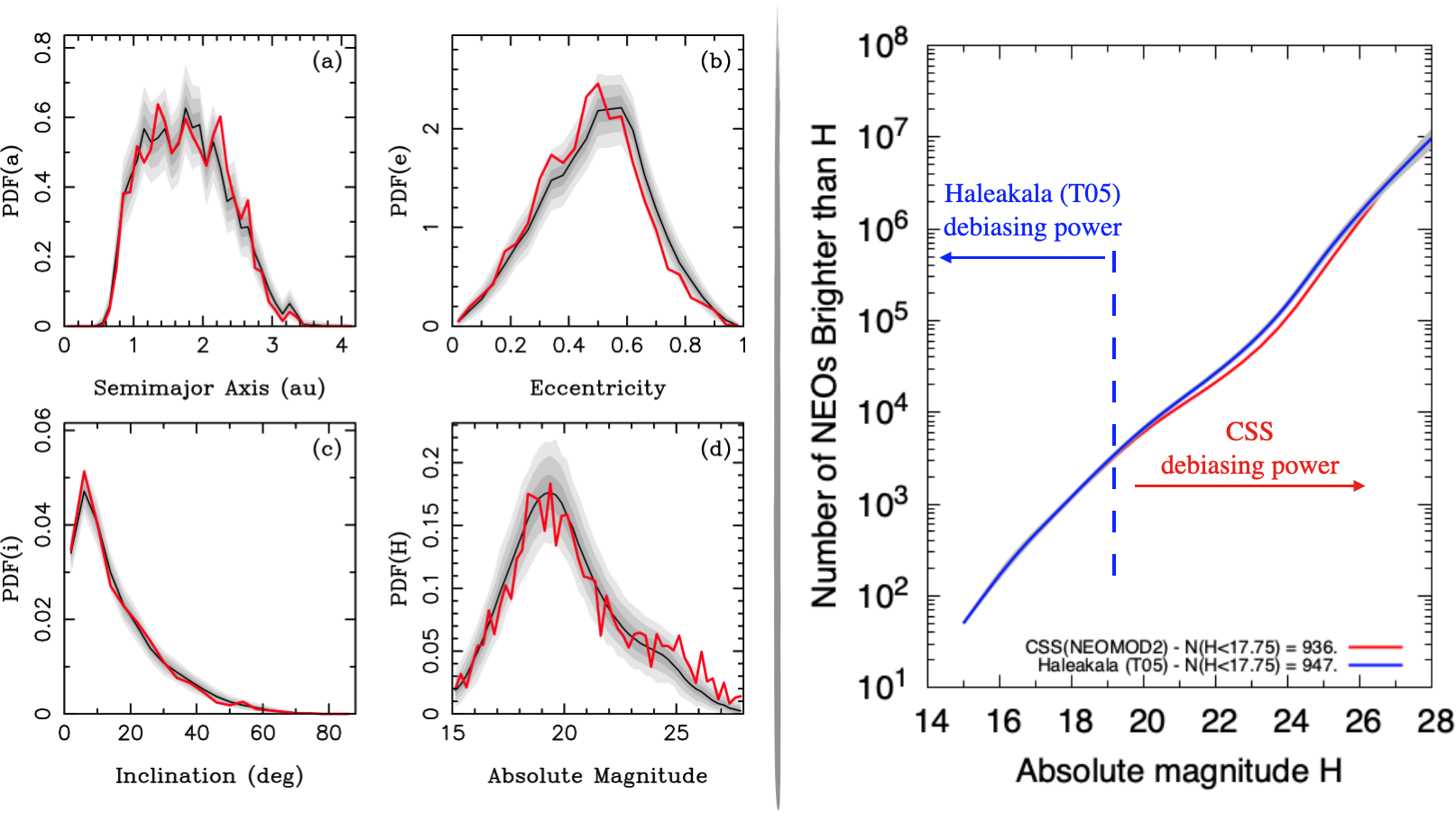}
    \caption{The same as in Figure \ref{Fig7} but for Haleakal\=a (T05 -- blue in the right panel) while considering a NEO sample of 3402.}
    \label{Fig8}
\end{figure}

We now perform similar analysis over Haleakela (T05). Figure \ref{Fig8} shows our comparison between our model biased PDFs and Haleakal\=a (T05)'s PDFs (four panels -- a, b, c, and d -- on the left side of the gray vertical line). Figure \ref{Fig8} (rightmost panel) also reports on the expected cumulative distribution of NEOs brighter than H compared to that from NEOMOD2 \citep{Nesvorny2024neomod2}. 

Haleakal\=a (T05) unique detection PDFs are closely represented by our model biased PDFs obtained via {\tt MultiNest} best-fit models. The synodic effects (`peaks and dips') from NEOs avoiding Haleakal\=a (T05)'s FoVs \citep[Figure \ref{Fig5} top right; see][]{Tricarico2017,Nesvorny2023neomod,Nesvorny2024neomod2}, as well as all other distributions, are recovered by our model biased PDFs within a 3$\sigma$ envelope. This implies our modeling is working reasonably well at debiasing this survey.

Haleakal\=a (T05) has its highest debiasing power of the NEO population for absolute magnitudes H $\lesssim$ 19 (dashed vertical blue line in the rightmost panel of Figure \ref{Fig8}; see also Figure \ref{Fig6}). In this magnitude range Haleakal\=a (T05) and CSS \citep[NEOMOD2;][]{Nesvorny2024neomod2} are in good agreement. The reference cumulative number of objects with H $<$ 17.75 according to Haleakal\=a (T05)'s debiased population is 947$^{+45}_{-45}$. Following discussion from previous Section, this suggests $\approx$ 90\%$^{+4\%}_{-4\%}$ population completeness, which, once again within uncertainties, is as complete as suggested by NEOMOD2 \citep[i.e., 91\%$^{+4\%}_{-4\%}$; Table 3 in][]{Nesvorny2024neomod2}.

We once again do not make any detailed comparison between Haleakal\=a (T05) and CSS for H $>$ 19 where CSS is expected to be more powerful at debiasing the NEO population. Nonetheless, we once again call attention to the fact that our debiased population estimation is similar to predictions made by NEOMOD2, despite the large difference in the number of unique detections from these two surveys in this range. CSS (after G96 camera update) had 9322 unique detections of NEOs with 19 $<$ H $<$ 28 while Haleakal\=a (T05) only 2196 in the same absolute magnitude range. 

\newpage
\subsection{Sutherland (M22)} \label{sec:atlas03a}

\begin{figure}[h!]
    \centering
    \includegraphics[width=1.\linewidth]{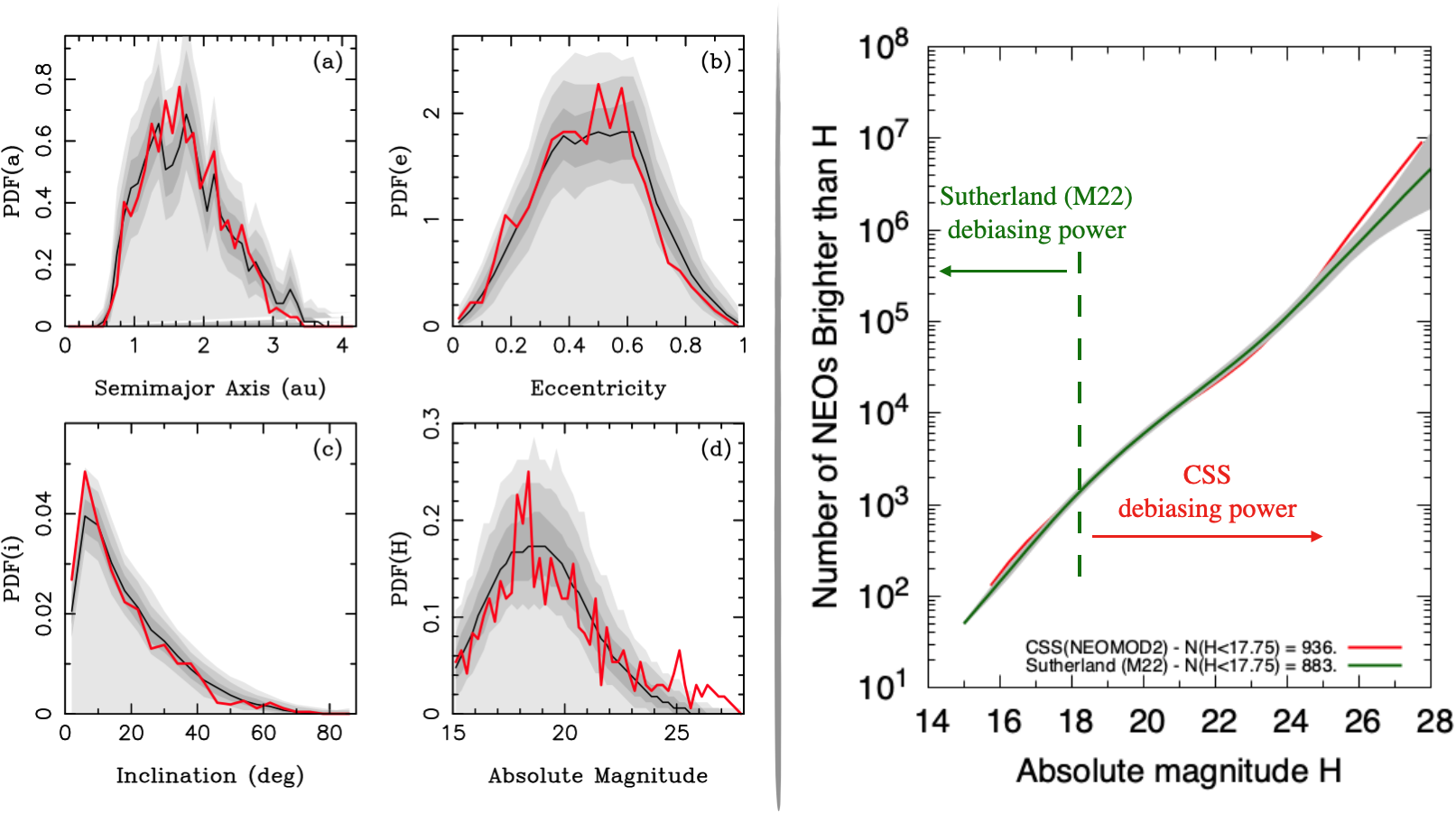}
    \caption{The same as in Figure \ref{Fig7} but for Sutherland (M22 -- green in the right panel) while considering a NEO sample of 671.}
    \label{Fig9}
\end{figure}

Moving to the southern telescopes of ATLAS, we now analyze how Sutherland (M22)'s model biased PDFs and debiased H distributions compare with the real detection PDF and CSS/NEOMOD2 predictions. Our results are presented in Figure \ref{Fig9}.

Our biased model PDFs are able of reproducing the overall shape of the real detection PDFs made by Sutherland (M22). However, we should note that in the case of this survey, our model bias PDFs have very large 3$\sigma$ envelopes. This is due to the fact that Sutherland (M22), which was only operational for one year  by the time we performed this work, had only 671 unique detections in the absolute magnitude range 15 $<$ H $<$ 28 (Figure \ref{Fig9}). These low number statistics create large fluctuations in the real detection PDFs. See, for instance, the large `peaks' in panel (d -- red curve in Figure \ref{Fig9}) around 17.5 $\lesssim$ H $\lesssim$ 19.5. Such large variations in the real data certainly affects the ability of {\tt MultiNest} to fit cubic splines \citep[Section \ref{sec:model}; see][for details on splines]{Nesvorny2023neomod} through the data. The coarse data, along with the very low amount of unique detections made by Sutherland (M22) in the range 25 $<$ H $<$ 28 (only 43 NEOs), also yields PDF(H) functions that fall short of reproducing the real detection PDF(H) in this range.    

The limited number of unique detections from Sutherland (M22) also interferes with the precision of {\tt MultiNest} in reproducing the `peaks and dips' related to synodic effects (panel a in Figure \ref{Fig9}), especially in the semimajor axis range 1.3 $\lesssim a \lesssim$ 1.7 au. Nonetheless, we notice that our model PDF(a) is more closely aligned to semimajor axes distances related to multiples of Earth's orbital period than the real data PDF (compare black and red in Figure \ref{Fig9} panel a). We argue that our model PDF provides a good representation of the expected semimajor axis distribution that Sutherland (M22) should be seeing if more detections were made. We reach this conclusion based on how synodic effects manifest themselves in the detection probability of both Mauna Loa (T08) and Haleakal\=a (T05). Both have one order of magnitude more unique detections (see top panels on Figure \ref{Fig5} and also panel (a) in Figures \ref{Fig7} and \ref{Fig8}) when compared to the lack of similar structures observed in the Figure \ref{Fig5} bottom left panel for Sutherland (M22)'s averaged detection probability. Therefore, such large 3$\sigma$ envelopes and coarse data should not be taken in a negative light. We consider our modeling of Sutherland (M22) to be good enough to evaluate what the debiased NEO population should be based on this survey.

The expected debiased and cumulative number of NEOs with H $<$ 17.75 by Sutherland (M22) is 883$^{+78}_{-80}$ (green in Figure \ref{Fig9} rightmost panel). This number of NEOs with H $<$ 17.75 suggests this population of NEOs should be complete to the 96\%$^{+10\%}_{-8\%}$ level, which is at face value $\approx$5\% more complete than what expected from NEOMOD2 \citep[i.e., 91\%$^{+4\%}_{-4\%}$; Table 3 in][]{Nesvorny2024neomod2}, even though uncertainties overlap. In Figure \ref{Fig9} (rightmost panel), we also used a vertical dashed green line to indicate the absolute magnitude where Sutherland (M22) debiasing power of the NEO population is higher (H $\lesssim$ 18). However, given the discussion from previous paragraphs in respect to low statistics, we prefer to refer to this value as a reference mark. 

The number of CSS unique detections (after G96 camera update) for 18 $<$ H $<$ 28 is 10,214 while Sutherland (M22) detected only 462 NEOs in the same absolute magnitude range. Yet, despite the large variation on the estimated number of NEO with H $>$ 25 (gray in Figure \ref{Fig9} rightmost panel -- associated to the discrepancy between biased model and real data PDF(H) -- Figure \ref{Fig9} panel d), our estimated debiased number of NEOs brighter than H from Sutherland (M22) is in good agreement with CSS estimates \citep[NEOMOD2;][]{Nesvorny2024neomod2}. This confirms our previous claim that our modeling of Sutherland (M22) is accurate enough for debiasing and understanding this survey.

\subsection{Chile (W68)} \label{sec:atlas04a}

\begin{figure}[h!]
    \centering
    \includegraphics[width=1.\linewidth]{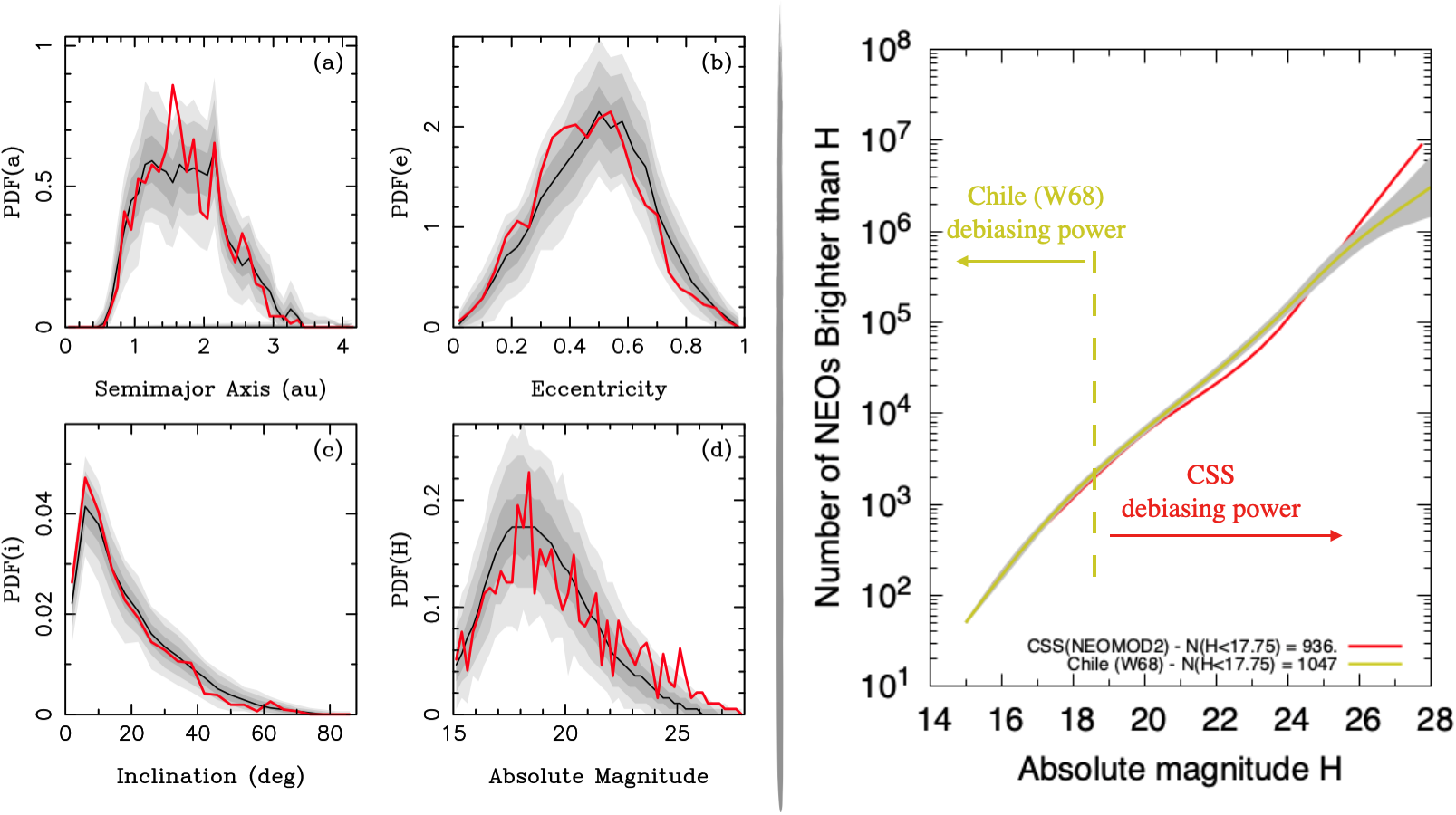}
    \caption{The same as in Figure \ref{Fig7} but for Chile (W68 -- yellow in the right panel) while considering a NEO sample of 779.}
    \label{Fig10}
\end{figure}

The last ATLAS telescope we analyze is Chile (W68). Figure \ref{Fig10} shows similar results to those in Figures \ref{Fig7}, \ref{Fig8}, and more specifically \ref{Fig9}. Results and conclusions from ATLAS Chile (W68) are very similar from Sutherland (M22), and they also `suffer' from low number statistics. Chile (W68) only had  779 unique detections within 15 $<$ H $<$ 28 (Figure \ref{Fig10}) over the period of performance we worked with. Nonetheless, as for Sutherland (M22), we can claim our modeling of Chile (W68) is accurate enough for debiasing and understanding this survey. 

The model biased PDFs are able to reproduce the overall shape of the real detection PDFs within 3$\sigma$ envelopes. Low statistics mostly affects Chile (W68) semimajor axis and absolute magnitude PDFs(a,H) in similar ways to Sutherland (M22). Yet, the Chile (W68) real detection PDF(a) seems coarse enough that {\tt MultiNest} struggles to fit the larger `peak' around $a\approx$ 1.5 au within 3$\sigma$.

The debiased cumulative H distribution of NEOs is in close agreement with CSS \citep[NEOMOD2;][]{Nesvorny2024neomod2} for H $\lesssim$ 19 (yellow vertical dashed line in Figure \ref{Fig10} rightmost panel). This is where Chile (W68) debaising power of the NEO population is the highest, according to our analysis from Section \ref{sec:performance} (see Figure \ref{Fig6}). Still, we prefer to simply refer to this value as a reference mark for Chile (W68). The estimated debiased number of NEOs with H $<$ 17.75 by Chile (W68) is 1047$^{+60}_{-69}$. This suggests $\approx$ 81\%$^{+6\%}_{-4\%}$ completeness of the population, about 10\% less than what was estimated by CSS \citep[NEOMOD2;][i.e. 91\%$^{+4\%}_{-4\%}$]{Nesvorny2024neomod2}. For absolute magnitudes between 19 $<$ H $<$ 28, the agreement between the debiased cumulative H distributions from Chile (W68) and CSS is not very good. This differs from the results shown for the other three ATLAS telescopes over this H range. Chile (W68) made 403 unique detections of NEOs with absolute magnitudes between 19 $<$ H $<$ 28, while CSS (after G96 camera update) made 9322 detections in the same H range. The low number of Chile (W68)'s unique detections in this H range, along with larger discrepancies in our model biased PDFs(a,H) may explain such differences. 

\subsection{ATLAS Debiased NEO Population} \label{sec:atlasall}

The final ATLAS model fit combined all four telescopes using the {\tt MultiNest} code (Section \ref{sec:model}). The goal was to determine the orbital distribution of NEOs from the best-fit (i.e., highest-likelihood) intrinsic model $\mathcal{M}$ (Eq.~\ref{eq4}). The code computed the log-likelihoods of each survey separately and combined them in order to evaluate the total log-likelihood (Eq.~\ref{eq3}). We argue that bringing together the surveys at the level of log-likelihoods is a better choice than trying to access their combined detection probabilities ($\mathcal{P}$) and number of unique detections. The reason is because the biased model ($\mathcal{M}_b$; Eq.~\ref{eq2}) of individual ATLAS telescopes, like Mauna Loa (T08) for instance, only applies to NEO detections made by the same individual telescope and not for any of the other three telescopes, and vice versa. Our method, differently than testing each survey separately makes use of the full statistical power of all four of them combined.

We model a total of 30 parameters, while using uniform priors \cite[Table \ref{Tab1}; see][for further details]{Nesvorny2023neomod,Nesvorny2024neomod2}. There are 12 magnitude-dependent weights $\alpha_j$, one for each of our NEO $n_s=$ 12 sources, i.e., eight individual resonances ($\nu_6$, 3:1, 5:2, 7:3, 8:3, 9:4, 11:5, and 2:1), inner main belt weak resonances, Hungarias and Phocaeas (representing high-inclination sources), as well as Jupiter-family comets (Section \ref{sec:model}). For target absolute magnitude values, {\tt MultiNest} evaluates $n_s-$1 values of $\alpha_j$ weights for those sources. The contribution by comets is obtained from $\Sigma_{j=1}^{n_s} \alpha_j$ = 1. We evaluate these magnitude-dependent weights at H = 15 and H = 28, expressed in Table \ref{Tab1} as $\alpha_j$'s for H = 15 with 1 $\leq j \leq$ 11 and $\alpha$'s for H = 28  with 12 $\leq j \leq$ 22. Our magnitude distribution parameters (H distribution in Table \ref{Tab1}) are defined by $N_{\rm ref}$ (the reference number of NEOs with H $<$ 17.75) and $\gamma_j$ (the slopes of the cumulative distribution N(H)), with 2 $\leq j \leq$ 6 (15 $\leq H \leq$ 28). The $\gamma_1$ parameter is fixed such that N(15) = 50. Finally, there are two additional parameters related to perihelion disruption model \citep[e.g.][]{Granvik2016}, i.e. $q_0^*$ and $\delta q^*$.

\begin{table}[]
    \centering
    \begin{tabular}{cccccc}
\hline
Label & Parameter & Median & -$\sigma$ & +$\sigma$ & Limit \\
\hline
$\alpha$'s for H = 15 & & & & & \\
(1) &  $\nu_6$  &  0.021  &  0.015  &  0.024  &  -  \\
(2) &  3:1  &  0.260  &  0.028  &  0.029  &  -  \\
(3) &  5:2  &  0.056  &  0.017  &  0.017  &  -  \\
(4) &  7:3  &  0.008  &  0.005  &  0.008  &  0.011  \\
(5) &  8:3  &  0.121  &  0.015  &  0.015  &  -  \\
(6) &  9:4  &  0.010  &  0.007  &  0.012  &  0.015  \\
(7) &  11:5  &  0.073  &  0.014  &  0.015  &  -  \\
(8) &  2:1  &  0.043  &  0.008  &  0.007  &  -  \\
(9) &  Inner weak  &  0.186  &  0.026  &  0.023  &  -  \\
(10) &  Hungarias  &  0.108  &  0.012  &  0.012  &  -  \\
(11) &  Phoccaeas  &  0.102  &  0.010  &  0.010  &  -  \\
(12) &  JFCs  &  0.005  &  0.003  &  0.004  &  0.007  \\
\hline
$\alpha$'s for H = 28 & & & & & \\
(12) &  $\nu_6$  &  0.551  &  0.046  &  0.043  &  -  \\
(13) &  3:1  &  0.216  &  0.038  &  0.038  &  -  \\
(14) &  5:2  &  0.017  &  0.011  &  0.017  &  0.024  \\
(15) &  7:3  &  0.005  &  0.004  &  0.008  &  0.009  \\
(16) &  8:3  &  0.015  &  0.011  &  0.020  &  0.023  \\
(17) &  9:4  &  0.008  &  0.006  &  0.012  &  0.012  \\
(18) &  11:5  &  0.008  &  0.006  &  0.013  &  0.013  \\
(19) &  2:1  &  0.011  &  0.008  &  0.014  &  0.018  \\
(20) &  Inner weak  &  0.111  &  0.048  &  0.053  &  -  \\
(21) &  Hungarias  &  0.020  &  0.014  &  0.024  &  0.030  \\
(22) &  Phoccaeas  &  0.006  &  0.005  &  0.009  &  0.010  \\
(23) &  JFCs  &  0.004  &  0.003  &  0.006  &  0.007  \\
\hline
H distribution & & & & & \\
(23) &  $N_{\rm ref}$  &  948  &  20  &  20  &  -  \\
(24) &  $\gamma_2$  &  0.422  &  0.010  &  0.010  &  -  \\
(25) &  $\gamma_3$  &  0.374  &  0.004  &  0.004  &  -  \\
(26) &  $\gamma_4$  &  0.332  &  0.003  &  0.003  &  -  \\
(27) &  $\gamma_5$  &  0.507  &  0.010  &  0.009  &  -  \\
(28) &  $\gamma_6$  &  0.412  &  0.015  &  0.015  &  -  \\
\hline
Disruption parameters & & & & & \\
(29) &  $q^*_0$  &  0.135  &  0.002  &  0.002  &  -  \\
(30) &  $\delta q^*$  &  0.116  &  0.006  &  0.005  &  -  \\
\hline
    \end{tabular}
    \caption{The median and uncertainties of our base model parameters. The uncertainties reported here were obtained from the posterior distribution produced by {\tt MultiNest}. They do not account for uncertainties of any of the four ATLAS telescopes detection efficiencies. For parameters, for which the posterior distribution peaks near zero, the last column reports the upper limit (68.3\% of posteriors fall between zero and that limit).}
    \label{Tab1}
\end{table}

The values reported in Table \ref{Tab1}  are the posterior distribution of model parameters provided by {\tt MultiNest}. These values are reasonably consistent with those found by \citet[compare with their Table 2]{Nesvorny2024neomod2}. We see, for example, that the contribution of the $\nu_6$ source is marginal at H = 15, but very high at H = 28. This outcome is the same conclusion made by \citet{Nesvorny2024neomod2}. The only moderate difference between our modeling of the NEO population by ATLAS and that from CSS \citep[NEOMOD2;][]{Nesvorny2024neomod2} is related to the contribution of the 3:1 and Inner main belt weak resonances to the population of small NEOs (H = 28). \citet{Nesvorny2024neomod2} reported a contribution of 31.3\% by the 3:1 and only 0.8\% by the Inner weak resonances. Our predictions from ATLAS instead suggest that at faint magnitudes the contribution by the 3:1 and Inner weak resonances are about 21.6\% and 11.1\% respectively. Yet, the sum of the contribution of 3:1 and Inner weak resonances is nearly the same in both predictions by CSS \citep[NEOMOD2;][]{Nesvorny2024neomod2} and ATLAS, i.e, 32.1\% and 32.7\%. This happens because {\tt MultiNest} found  a different split between 3:1 and Inner weak resonances while keeping their overall contribution roughly the same. The likely reason is that there is some source degeneracy in predictions by ATLAS, CSS, or both. The differences between the results are small, though. 

\begin{figure}[h!]
    \centering
    \includegraphics[width=0.5\linewidth]{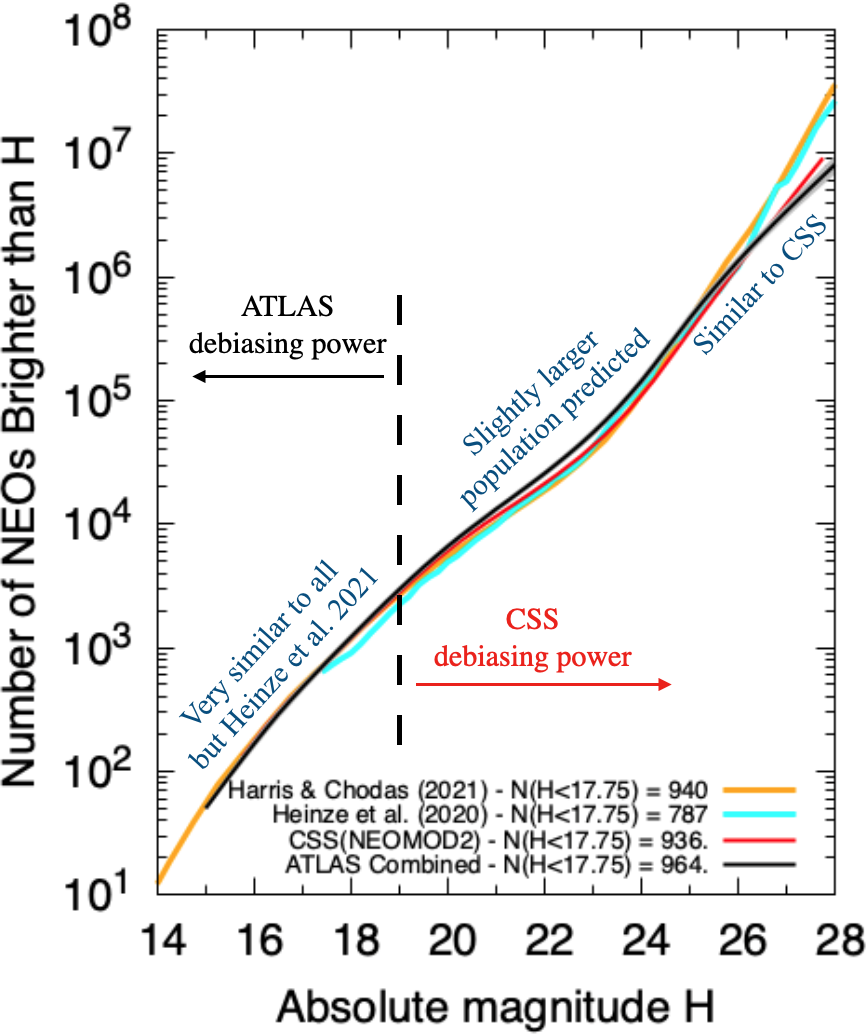}
    \caption{The intrinsic (debiased) absolute magnitude distribution of NEOs from our base model (black line is the median), estimated once combining all four ATLAS telescopes, is compared to the magnitude distribution from \citet[red line]{Nesvorny2024neomod2}, \citet[cyan line]{Heinze2021}, and \citet[orange line; see also \citet{Harris2023}]{Harris2021}. The gray area is the 3$\sigma$ envelope obtained from the posterior distribution computed by {\tt MultiNest}. It contains 99.7\% of our base model posteriors. 
    The vertical black dashed line indicates the approximate location of the H value where both ATLAS and CSS are expected to have similar debiasing power of the NEO population, according to what was reported in  Section \ref{sec:performance} (Figure \ref{Fig6}). Labels in the bottom of this plot report on the expected cumulative number of objects with H $<$ 17.75 (D $\approx$ 1 km for $p_V=$0.14) for ATLAS and from relevant literature. Labels in dark blue color give a brief description of main features of ATLAS fit when compared to others.}
    \label{Fig11}
\end{figure}

Figure \ref{Fig11} shows our intrinsic (bias-corrected) absolute magnitude distribution from our base model $\mathcal{M}$ accounting for the contribution of all ATLAS telescopes combined (black curve). This distribution is very similar to that reported by \citet[orange; see also \citet{Harris2023}]{Harris2021} and \citet[red]{Nesvorny2024neomod2} for H $<$ 19 (where ATLAS debiaing power of the NEO population is expected to be the highest - Section \ref{sec:performance}, Figure \ref{Fig6}), but it differs from \citet[cyan]{Heinze2021}. As reported in the label at the bottom right of Figure \ref{Fig11}, the reference number of NEOs with H $<$ 17.75 from our ATLAS fit is 964$_{\rm -29}^{\rm +27}$ (see also Table \ref{Tab2}), which is similar (within error bars) to the $\approx$ 940 estimated by \citet[see their Table in Appendix B]{Harris2021} and 936$_{\rm -38}^{\rm +41}$ by \citet[see their Table 3]{Nesvorny2024neomod2}. For further reference, our reported numbers are also in good agreement with those from \citet{Granvik2018}, i.e. 962$_{\rm -52}^{\rm +56}$. \citet{Heinze2021} estimated 787 objects with  
H $<$ 17.75 (D $\approx$ 1 km for $p_V=$ 0.14)\footnote{The number of NEOs brighter than H = 17.75 from \citet[see their Table 3]{Heinze2021} was computed by interpolating their reported values at N(17.6) = 718.9 and N(17.8) = 809.9. We do not attempt to interpolate their uncertainties.}. Our estimate of 964$_{\rm -29}^{\rm +27}$ NEOs with H $<$ 17.75 suggest this population completeness is roughly 88\%$_{\rm -2\%}^{\rm +3\%}$ (Table \ref{Tab2}), therefore, a few percent less complete than previously suggested by \citet[i.e. 95.5\%]{Harris2021} but still within the error bars from \citet[i.e. 91\%$_{\rm -4\%}^{\rm +4\%}$,]{Nesvorny2024neomod2}, whose values were obtained directly from their works and are not recomputed here (see discussion in Section \ref{sec:atlas01a} related to estimates in completeness from different MPC catalogs). \citet{Heinze2021} do not report on their completeness. \citet[see their Table 1]{Grav2023} reported that NEOs brighter than H = 17.75 continue to be discovered at a steady rate of more than one object per year. Indeed, two additional discoveries were made in 2023 after February 3, 2023 (2023~HQ2 and 2023~GZ1 with H = 16.33 and 17.61 according to the MPC database as of February 20, 2024). Those new discoveries are not accounted for in our present work.

For H $>$ 19 (where ATLAS debiasing power of the NEO population decreases when compared to CSS - Section \ref{sec:performance}, Figure \ref{Fig6}) we find that our estimate of the number of NEOs brighter than H diverges slightly from \citet{Harris2021,Nesvorny2024neomod2}, and \citet{Heinze2021}. The number of NEOs larger than H predicted from ATLAS within the range 19 $<$ H $\lesssim$ 25 is slightly larger than from any other work \citep[yet with similar slopes $\gamma_4$ and $\gamma_5$ in this H distribution range, as those reported by][see their Table 2]{Nesvorny2024neomod2}. More specifically to objects brighter than H = 22.02 \citep[D $\approx$ 140 m for $p_V=$ 0.14; the target size to be addressed by the U.S. Congressional mandate; e.g.][]{Grav2023} we estimate N(22.02) $\approx$ 26,460\footnote{\citet{Granvik2018} estimate about 25,590$^{\rm +479}_{\rm -440}$ objects with H $<$ 22.}. The current number of known objects brighter than H = 22.02 from the MPC catalog as of February 3, 2023 is 10,401. This leads to an estimated completeness of about 39\% from ATLAS. 

\citet{Harris2021} and \citet{Nesvorny2024neomod2} do not report on their completeness estimate specifically for N(22.02), thus, as previously discussed earlier in this Section and in Section \ref{sec:atlas01a}, due to variations in H magnitude from updates in the MPC catalog over the years \citep[e.g.,][]{Pravec2012}, we cannot compare our estimate with their work directly. Yet, as the exact diameter value of 140 m strongly depends on the assumed value for geometric albedo \citep[which we do not know for most of the objects with absolute magnitudes in the range 21.75 $<$ H $<$ 22.25; see][]{Nesvorny2024neomod3}, in Table \ref{Tab2} we more broadly report on completeness for H $<$ 22.25 when targeting for the completeness of the population with D $\gtrsim$ 140 m objects. 

\citet{Harris2021} and \citet{Nesvorny2024neomod2} also reported on their estimated completeness for N(22.25), which allows us to make a comparison. Under these considerations, from ATLAS estimates, N(22.25) = 30,600$_{\rm -800}^{\rm +800}$, which suggests 36\%$_{\rm -1\%}^{\rm +1\%}$ completeness of the population of NEOs with D $\gtrsim$ 140 m. This indicates that according to ATLAS estimates, this population is about 8\%$_{\rm -1\%}^{\rm +1\%}$ less complete than the 44\% value reported by both \citet[Appendix B]{Harris2021} and \citet[Table 3]{Nesvorny2024neomod2}. Nonetheless, it is worthy remembering that CSS \citep[NEOMOD2;][]{Nesvorny2024neomod2} should be more precise in debiasing the NEO population at H $>$ 19 than ATLAS as indicated in Figure \ref{Fig11} (see also Section \ref{sec:performance} and Figure \ref{Fig6}). Finally, \citet{Nesvorny2024neomod3}, while including the full range of albedos from WISE on NEOMOD3 (see Section \ref{sec:intro}), reports on 20,000$_{\rm -2000}^{\rm +2000}$ NEOs larger than D = 140 m.

For H $>$ 25 our intrinsic (debiased) model $\mathcal{M}$ is very similar to that reported by \citet{Nesvorny2024neomod2}. Our H distribution between 25 $<$ H $<$ 28 is much shallower than \citet{Harris2021} and \citet{Heinze2021}. The differences between our distribution and that by \citet{Heinze2021} is expected to be due to the differences in our considerations regarding trailing losses (Section \ref{sec:trlloss}). Due to the fact that ATLAS is not optimal in debiasing such faint populations (recall that our model biased PDFs fall a bit short in reproducing PDF(H) for H $>$ 25 in panel d from Figures \ref{Fig7} to \ref{Fig10}), we cannot make precise or strong conclusions in regards to why our results are different from those by \citet[see also \citet{Harris2023}]{Harris2021}. Yet, similarities between our H distribution with that predicted from CSS \citep[NEOMOD2;][]{Nesvorny2024neomod2} may lead to the same conclusions made by those authors. 

%
\begin{table}[]
    \centering
    \begin{tabular}{cccccccccccc}
\hline
$H_1$ & $H_2$ & d$N$ & $N(H_2)$ & $N_{\rm HC}(H_2)$ & $N_{\rm H21}(H_2)$ & $N_{\rm CSS}(H_2)$ & $N_{\rm min}(H_2)$ & $N_{\rm max}(H_2)$ & $N_{\rm MPC}(H_2)$ & Compl. & Range \\
\hline
15.25  &  15.75  &  56  &  124  &  136  &  -  &  130  &  120  &  128  &  122  &  98\% &  (95-100)  \\
15.75  &  16.25  &  96  &  220  &  235  &  -  &  234  &  212  &  230  &  210  &  95\% &  (91-99)  \\
16.25  &  16.75  &  152  &  373  &  398  &  -  &  390  &  358  &  388  &  360  &  97\% &  (93-100)  \\
16.75  &  17.25  &  232  &  605  &  621  &  -  &  608  &  586  &  624  &  558  &  92\% &  (89-95)  \\
17.25  &  17.75  &  359  &  964  &  940  &  787  &  936  &  935  &  991  &  850  &  88\% &  (86-91)  \\
17.75  &  18.25  &  561  &  1520  &  1474  &  1106  &  1470  &  1480  &  1560  &  1324  &  87\% &  (85-89)  \\
18.25  &  18.75  &  853  &  2380  &  2210  &  1807  &  2240  &  2320  &  2430  &  2026  &  85\% &  (83-87)  \\
18.75  &  19.25  &  1267  &  3640  &  3230  &  2675  &  3410  &  3570  &  3730  &  2901  &  80\% &  (78-81)  \\
19.25  &  19.75  &  1802  &  5440  &  4625  &  4001  &  5050  &  5330  &  5570  &  4025  &  74\% &  (72-76)  \\
19.75  &  20.25  &  2450  &  7890  &  6419  &  5668  &  7210  &  7710  &  8070  &  5304  &  67\% &  (66-69)  \\
20.25  &  20.75  &  3250  &  11100  &  8731  &  8277  &  9920  &  10900  &  11400  &  6685  &  60\% &  (59-61)  \\
20.75  &  21.25  &  4370  &  15500  &  11768  &  11853  &  13400  &  15200  &  15800  &  8156  &  53\% &  (52-54)  \\
21.25  &  21.75  &  6110  &  21600  &  15880  &  16680  &  18100  &  21100  &  22100  &  9606  &  44\% &  (43-46)  \\
21.75  &  22.25  &  8960  &  30600  &  21681  &  24023  &  24900  &  29800  &  31400  &  11032  &  36\% &  (35-37)  \\
22.25  &  22.75  &  14010  &  44600  &  31431  &  32730  &  35300  &  43300  &  45900  &  12532  &  28\% &  (27-29)  \\
22.75  &  23.25  &  23300  &  67900  &  47577  &  55745  &  52500  &  65800  &  70100  &  14133  &  21\% &  -  \\
23.25  &  23.75  &  41800  &  110000  &  82556  &  99650  &  83600  &  106000  &  113000  &  16006  &  15\% &  -  \\
23.75  &  24.25  &  80000  &  190000  &  153000  &  166425  &  144000  &  182000  &  197000  &  18037  &  9.5\% &  -  \\
24.25  &  24.75  &  153700  &  343000  &  313000  &  317675  &  266000  &  328000  &  359000  &  20229  &  5.9\% &  -  \\
24.75  &  25.25  &  268000  &  612000  &  641000  &  562825  &  494000  &  581000  &  644000  &  22356  &  3.7\% &  -  \\
25.25  &  25.75  &  427000  &  1040000  &  1300000  &  927275  &  905000  &  986000  &  1100000  &  24362  &  2.3\% &  -  \\
25.75  &  26.25  &  666000  &  1710000  &  2410000  &  1806500  &  1630000  &  1610000  &  1810000  &  26159  &  1.5\% &  -  \\
26.25  &  26.75  &  1013000  &  2720000  &  4810000  &  4908500  &  2920000  &  2530000  &  2900000  &  27645  &  1.0\% &  -  \\
26.75  &  27.25  &  1517000  &  4230000  &  10800000  &  8393000  &  5170000  &  3870000  &  4590000  &  28817  &  0.7\% &  -  \\
27.25  &  27.75  &  2290000  &  6520000  &  24400000  &  19392500  &  9120000  &  5810000  &  7270000  &  29668  &  0.5\% &  -  \\
\hline
    \end{tabular}
    \caption{The absolute magnitude distribution and completeness of the NEO population. The columns are: the lower limit of a magnitude bin (H$_{\rm 1}$), upper limit of a magnitude bin (H$_{\rm 2}$), ATLAS estimate of the number of NEOs between H$_{\rm 1}$ and H$_{\rm 2}$ (dN), ATLAS estimate of the number of NEOs with N $<$ H$_{\rm 2}$ (N(H$_{\rm 2}$)), \citet{Harris2021} estimate of
N(H$_{\rm 2}$) (N$_{\rm HC}$(H$_{\rm 2}$)), \citet{Heinze2021} estimate of
N(H$_{\rm 2}$) (N$_{\rm H21}$(H$_{\rm 2}$), where we interpolated between the limits provided in their Table 3 as our reported H intervals differ from each other), \citet[NEOMOD2]{Nesvorny2024neomod2} estimate of N(H$_{\rm 2}$) (N$_{\rm CSS}$(H$_{\rm 2}$)), ATLAS estimate of N(H$_{\rm 2}$) minus 1$\sigma$ (N$_{\rm min}$(H$_{\rm 2}$)), ATLAS estimate of N(H$_{\rm 2}$) plus 1$\sigma$ (N$_{\rm max}$(H$_{\rm 2}$)), number of NEOs with N $<$ H$_{\rm 2}$ in the MPC catalog as of February 3, 2023 (N$_{\rm MPC}$(H$_{\rm 2}$)), completeness defined as N$_{\rm MPC}$(H$_{\rm 2}$)/N(H$_{\rm 2}$), and 1$\sigma$ completeness range ($<$1\% uncertainties not listed).
}
    \label{Tab2}
\end{table}

\newpage
\section{Summary/Conclusions} \label{sec:conclusion}

The main results of this work are summarized as follow:

\begin{enumerate}
    \item We debiased the population of NEOs primarily based on observations from each individual ATLAS telescopes, i.e., Mauna Loa (T08), Haleakal\=a (T05), Sutherland (M22), and Chile (W68); see Figures \ref{Fig7} to \ref{Fig10}. We then combined all four telescopes log-likelihoods from our {\tt MultiNest} best fits to access the overall NEO population absolute magnitude distribution and their completeness. Our results show good agreement with previous works \citep[i.e., NEOMOD2][see Figure \ref{Fig11} and Tables \ref{Tab1} and \ref{Tab2}.]{Nesvorny2024neomod3}.
    
    \item Our median fit values reported in Table \ref{Tab1} are similar to those from Table 2 in \citet{Nesvorny2024neomod2}. This emphasizes the $\nu_6$ secular resonance is the largest source of contribution for the small and faint NEO population at H = 28, and with little contribution to bright NEOs at H = 15, where the 3:1 MMR contribution is larger. We also reported on the possibility of source degeneracy between 3:1 and Weak Inner resonances for faint NEOs at H = 28 for both ATLAS and CSS. 

    \item When comparing ATLAS and CSS debiasing power of the NEO population, we find that ATLAS is more reliable at debiasing the NEO population at H $\lesssim$ 19 mag (Figure \ref{Fig6}). This is mostly because ATLAS has a larger sky coverage than CSS, which allows ATLAS to observe portions of the sky that are not covered by CSS observations (see Figures \ref{Fig1} and \ref{Fig2}, as well as discussion related in Section \ref{sec:intro}), thus allowing ATLAS to find bright NEOs that would be constantly hiding from CSS. In this regards, ATLAS Southern telescopes Sutherland (M22) and Chile (W68) are important additions to the ATLAS network due to their constant monitoring of the not well observed (by CSS - Figure \ref{Fig1} Panel A) south hemisphere sky, where for instance, 2022~RX3 (H = 17.7) was discovered by Chile (W68 -- Figure \ref{Fig2}). 

    \item Our debiased cumulative magnitude distribution at H $\lesssim$ 19 (Figure \ref{Fig11}) shows good agreement with previous estimates by \citet[see also \citet{Harris2023}]{Harris2021} and \citet{Nesvorny2024neomod2}, while improving over the finds by \citet{Heinze2021}. ATLAS estimate of the NEO population with cumulative magnitude distribution in the range 19 $\lesssim$ H $\lesssim$ 25 is slightly larger than those from \citet{Harris2021}, \citet{Nesvorny2024neomod2}, and \citet{Heinze2021}. Yet, the number of objects with D $<$ 140 m (H $\approx$ 22.02 assuming fixed albedo $p_V=$ 0.14) from ATLAS ($\approx$ 26,460) is close to the estimate from \citet[NEOMOD3]{Nesvorny2024neomod3} when using WISE albedo distribution combined with CSS observations (NEOMOD2), i.e., N($<$D = 140 m) = 20,000$\pm$2000. While perhaps not optimal for debiasing the NEO population with H $\gtrsim$ 25, ATLAS estimate of that cumulative population is similar to that by CSS \citep[NEOMOD2][]{Nesvorny2024neomod2}, implying similar conclusions to that work.

    \item Finally we estimate the NEO population completion to be $\approx$ 88\%$^{+3\%}_{-2\%}$ for NEOs with H $<$ 17.75 and $\approx$ 36\%$^{+1\%}_{-1\%}$ for NEOs with H $<$ 22.25 (Table \ref{Tab2}). Our completeness estimate for the NEO population with H $<$ 17.75 are within the uncertainties reported by \citet[i.e., 91\%$^{\rm +4\%}_{\rm -4\%}$]{Nesvorny2024neomod2}, which used similar debiasing methods, but are some 7\%$^{+2\%}_{-3\%}$ lower than that reported by \citet[i.e., $\approx$95\%]{Harris2021,Harris2023}, which relied on re-detections for debiasing the NEO population. Our estimate of the NEO population completeness for H $<$ 22.25 on the other hand is about 8\%$^{\rm +1\%}_{\rm -1\%}$ lower than what reported by both \citet{Harris2021} and \citet{Nesvorny2024neomod2}. ATLAS has a high debiasing power for H $\lesssim$ 19 (Figure \ref{Fig6}), as well as  an all-sky coverage that gives it the ability of looking at regions of the sky (e.g., Southern hemisphere) and finding bright NEOs at geometries that other surveys cannot (e.g., CSS -- see Figure \ref{Fig1} and \ref{Fig2}). Yet, similarly to \citet{Nesvorny2024neomod2}, we must acknowledge that, given the number of NEOs with H $<$ 17.75 is relatively small, accounting for re-detections \citep{Harris2021,Harris2023} may improve statistics and potentially accuracy in debiasing such population of bright NEOs over methods that rely on unique detections. Nonetheless, we find the reported differences to be overall small, and that it is noteworthy the constancy of the results by multiple studies using different models, telescope systems, and assumptions. This indicates a great stability in the results and that the overall literature related to evaluating the NEO population and surveys completion is moving in the right direction. 
    
\end{enumerate}

\section*{Acknowledgements}
We thank Alan W. Harris and Mikael Granvik for their constructive suggestions that helped improving this work.
The simulations were performed on the NASA Pleiades Supercomputer. We thank the NASA NAS computing division for continued support. The work of RD, DN, and WFB was supported by the NASA Planetary Defense Coordination Office project ``Constructing a New Model of the Near-Earth Object Population''.
The work of DV was partially supported by the Czech Science Foundation (project No. 23-04946S). The work of SN, SRC, PWC, and DF was conducted at the Jet Propulsion Laboratory, California Institute of Technology, under a contract with the National Aeronautics and Space Administration. This work has made use of data from the Asteroid Terrestrial-impact Last Alert System (ATLAS) project. ATLAS is primarily funded to search for near earth asteroids through NASA grants NN12AR55G, 80NSSC18K0284, and 80NSSC18K1575; byproducts of the NEO search include images and catalogs from the survey area.  The ATLAS science products have been made possible through the contributions of the University of Hawaii Institute for Astronomy, the Queen's University Belfast, the Space Telescope Science Institute, the South African Astronomical Observatory (SAAO),  and the Millennium Institute of Astrophysics (MAS), Chile.


\newpage




\end{document}